\documentclass[nohyper,12pt,letterpaper]{JHEP3}
\usepackage[dvips]{epsfig}
\usepackage{amsfonts,amssymb}

\newcommand{\be}{\begin{equation}}
\newcommand{\ee}{\end{equation}}
\newcommand{\ben}{\begin{displaymath}}
\newcommand{\een}{\end{displaymath}}
\newcommand{\bea}{\begin{eqnarray}}
\newcommand{\eea}{\end{eqnarray}}
\newcommand{\bean}{\begin{eqnarray*}}
\newcommand{\eean}{\end{eqnarray*}}

\def\l {\lambda}
\def\a {\alpha}

\def\b {\beta}

\def\d {\delta}

\def\e {\epsilon}

\newcommand{\ads}[1]{\mbox{${AdS}_{#1}$}}
\newcommand{\adss}[2]{\mbox{$AdS_{#1}\times {S}^{#2}$}}

\newcommand{\ket}[1]{\mbox{$| #1 \rangle$}}

\newcommand{\ie}{{\it i.e.}}

\newcommand{\la}{\zeta}
\newcommand{\lap}{\zeta_+}
\newcommand{\lam}{\zeta_-}
\newcommand{\lapm}{\zeta_\pm}
\newcommand{\lab}{\bar{\la}}
\newcommand{\labm}{\bar{\la}_-}

\newcommand{\buo}{\hat{u}}

\newcommand{\rr}{r}

\newcommand{\thc}{\lambda}

\newcommand{\dmu}{\hat{\mu}_1}

\newcommand{\ekpl}{e^{ik_{P_l}}}
\newcommand{\ekplo}{e^{ik_{P_{l+1}}}}

\newcommand{\mathbc}{\mathbf}

\newcommand{\one}{\mathbc{1}}
\newcommand{\two}{\mathbc{2}}
\newcommand{\three}{\mathbc{3}}
\newcommand{\four}{\mathbc{4}}
\newcommand{\five}{\mathbc{5}}
\newcommand{\six}{\mathbc{6}}

\newcommand{\Ima}{\mathfrak{Im}}
\newcommand{\Rea}{\mathfrak{Re}}

\newcommand{\beq}{\begin{equation}}
\newcommand{\eeq}{\end{equation}}
\newcommand{\beqr}{\begin{displaymath}}
\newcommand{\eeqr}{\end{displaymath}}
\newcommand{\beqa}{\begin{eqnarray}}
\newcommand{\eeqa}{\end{eqnarray}}
\newcommand{\beqar}{\begin{eqnarray*}}
\newcommand{\eeqar}{\end{eqnarray*}}
\renewcommand{\k}{\kappa}
\newcommand{\cN}{{\cal N}}

\newcommand{\cL}{{\cal L}}
\newcommand{\cW}{{\cal W}}

\newcommand{\half}{\ensuremath{\frac{1}{2}}}

\newcommand{\N}[1]{\ensuremath{\cN=#1}}

\def\ci{\cite}
\def\ov{\over }

\title{\LARGE
 Spiky strings  and  giant magnons on $S^5$
}

\author{

M. Kruczenski$^{1,}$\footnote{After August 15$^{th}$ at: Physics Department,
Purdue University, 525 Northwestern Avenue, W. Lafayette, IN
47906-2036, USA.}, J. Russo$^{2}$
and A.A. Tseytlin$^{3,}$\footnote{Also at the Ohio State University
and  Lebedev
 Institute, Moscow.}\\
\vskip 0.4truecm
$^{1}$
{\it Physics Department, Princeton University, \\ Princeton, NJ 08540, USA.}
\vskip .2truecm
$^{2}$ {\it
Instituci\' o Catalana de Recerca i Estudis Avan\c{c}ats (ICREA),\\
Departament ECM,
Facultat de F\'\i sica, Universitat de Barcelona,
 Spain}
\vskip .2truecm
$^{3}$ {\it Blackett Laboratory,
Imperial College,\\
London, SW7 2AZ, UK}\\

E-mail: \email{martink@princeton.edu, jrusso@ub.edu, tseytlin.1@osu.edu}
}

\abstract{Recently, classical solutions for strings moving in
\adss{5}{5} have played an important role in understanding
the AdS/CFT correspondence. A large set of them were shown to follow
from an ansatz that reduces the solution of the string  equations
of motion to the study of a well-known integrable 1-d system known as
the Neumann-Rosochatius (NR) system. However,  other simple solutions such
as spiky strings or giant magnons in $S^5$   were not included in the
NR ansatz. We show that, when considered in the conformal gauge, these
solutions can be also accomodated by a version  of the NR-system.
This allows us to   describe
in detail a giant magnon solution with two additional angular momenta
and show that it can be interpreted
 as a superposition of two magnons moving with the same speed. 
In addition, we  consider  the spin chain side   and  describe
the corresponding state as that of two bound states in the infinite $SU(3)$ spin 
chain. We construct the Bethe ansatz wave function for such bound
state.
}

\keywords{spin chains, string theory, AdS/CFT}

\preprint{\tt{IMPERIAL-TP-AT-6-5}\\
          \tt{PUPT-2203}  }

\begin{document}

\section{Introduction}

 The AdS/CFT correspondence~\cite{malda} provided the first concrete example 
 of a large-$N$ duality\cite{largeN} between a gauge theory and a 
string theory in four dimensions. It is important to fully understand how  string theory emerges here from the  field theory since this  might later
provide methods applicable to other gauge theories. The basic 
 example is the relation between large-$N$, \N{4} super Yang-Mills (SYM) and IIB strings
in \adss{5}{5} \cite{malda}. In that case it is possible 
to see how certain simple  string states actually appear as field theory 
operators \cite{bmn,GKP} under the duality map. An important role is played, in particular,  by more general multi-spin  rotating string solutions on $S^5$ 
introduced  in \cite{FT}.
 The field theory description of such strings is in terms of semiclassical states of spin chains. The spin chain picture of the corresponding 
scalar field theory operators and their anomalous dimensions was found in \cite{MZ}, and the leading-order spin chain 
states corresponding to the 2-spin  rotating strings were found in \cite{BFST}. The semi-classical nature of these states
was emphasized in \cite{kru,KRT} where also a direct relation between the two low-energy effective field theory systems was described.    
 
While the integrability of the classical string sigma model implies  a general description of the (``finite-gap'') 
classical solutions in terms of solutions of certain integral equations
 \cite{KMMZ}, it is still important to 
find explicitly more generic yet simple string solutions and identify their corresponding duals. 
 In \cite{AFRT,ART} a generalized ansatz was proposed which reduces the problem of finding a large class of solutions to that 
of solving an integrable  one-dimensional system  -- Neumann  system,
 describing  an oscillator on a sphere. 
A particular reduction of the Neumann system leads to the  so called Neumann-Rosochatius (NR) system which describes a particle on a sphere 
subject to a sum of $r^2$  and $1 \over r^2$ potentials. This is again a  well-known integrable system whose integrals of motion 
and  solutions can be found rather explicitly \cite{NR,Moser}. The corresponding semiclassical solutions  correspond, in particular, 
 to folded, bended, wound rigid rotating strings  on $S^5$. One arrives at the NR action by choosing the conformal gauge 
and assuming  a particular ansatz for string coordinates (``NR-ansatz''). 

  However, some  other  important string configurations such as strings with spikes \cite{spiky,Ryang} and (bound states of) 
giant magnons \cite{HM,Dorey,Dorey2} (see also \cite{AFZ,MTT}) were not  found using an NR-type system.
They were first obtained  using  the Nambu-Goto action in the static-type gauge.\footnote{Conformal gauge was used also in \cite{Dorey2,AFZ};
their solution  for a giant magnon with spin  is equivalent to the one discussed below.}
    
Below we shall show that if one starts with   the conformal gauge, both the spiky strings 
 and the giant magnons can be  described  by a
  generalization of the NR ansatz of \cite{ART}. 
In this way it is possible to see that, in fact,  the giant magnon solutions (with additional spins)  
are a particular limit of the spiky solutions (the latter can, in turn,  be viewed as  superpositions of giant magnons). 
However, this is an important limit  since the solutions simplify substantially  when one of the three $S^5$ momenta is sent to infinity.
  

\bigskip

 The paper is organized as follows.
 In section 2 we shall introduce a generalized NR ansatz
 that describes solutions  with spikes and 3 angular momenta on $S^5$.
 Then in sections 3 and 4 we shall describe  solutions with two and three
 non-zero angular momenta. In  particular, we shall explicitly 
 present a generalization of the giant magnon  which
  carries two additional angular momenta and discuss the interpretation 
  of this new solution.
  In section 5 we shall consider in detail 
  the  dual spin chain  description 
  of the corresponding gauge theory states. 
  Some conclusions will be presented in section 6.

\section{Spiky strings and NR model }

We want to generalize the spiky solutions on $S^5$ to add more rotations and also make contact with giant magnons. 
The spiky solutions were originally  constructed  in \cite{spiky} as describing strings rotating in \ads{5} but here 
we are interested in generalizing their $S^5$  analog considered previously in \cite{Ryang}. The aim is to find them 
as solutions of an  NR-type  ansatz  similar to  the one in \cite{AFRT,ART}.

Let us start  with  the flat space string-with-spikes solution~\cite{flatc,spiky} which
is easily written in conformal gauge. If the flat metric on $R_t \times R^2$ is
\beq
ds^2 = -dt^2 + dXd\bar{X}
\eeq
then the  spiky solution  is  ($n$ is the number of spikes):
\beqa
 t = \tau \ , \ \ \ \ \ \  \   \ \ 
 X = e^{i(n-1)(\tau+\sigma)} + (n-1) e^{i(\tau-\sigma)} \ . 
\eeqa
Introducing the notation:
\beq
\omega= 2 \frac{n-1}{n}\ , \ \ \ \ \ \ \ \ \ \ \ \ \ \xi = \sigma+\frac{n-2}{n}\tau\ , 
\eeq
we  can write 
\beq
X = \left[e^{i(n-1)\xi}+(n-1)e^{-i\xi}\right] e^{i\omega\tau} =
\ x(\xi)\ e^{i\omega\tau} \ . 
\eeq
 This  looks similar to the ansatz in \cite{ART}  with spatial dependence 
 of the ``radial'' direction $x$  extended to dependence 
 on a linear combination of $\sigma $   and $\tau$.

\subsection{Generalized NR  ansatz }

Let us  now consider a string moving on an odd-dimensional sphere using conformal gauge. Then the metric is (in the $S^5$ case of interest $a=1,2,3$)
\beq
 ds^2 = -dt^2 + \sum_a\, dX_a d\bar{X}_a , \ \ \ \ \ \ \ \ \ \ \ 
  \sum_a |X_a|^2 =1  \ ,  
\eeq
so that the string Lagrangian becomes 
\beq
 \cL = -(\partial_\tau t)^2 + (\partial_\sigma t)^2 
 + \sum_a \left[\partial_\tau X_a \partial_\tau \bar{X}_a - \partial_\sigma X_a \partial_\sigma \bar{X}_a \right]
  - \Lambda \left( \sum_a X_a \bar{X}_a -1\right) \ . 
\label{orlag}
\eeq
whereas the action is:
\beq
S = \frac{T}{2} \int \cL
\eeq
According to the AdS/CFT correspondence, the string tension $T$ is a function of the 't Hooft coupling $\thc$ of the dual gauge theory: 
$T =\frac{\sqrt{\lambda}}{2\pi}$.  

 The equation of motion for $t$ is satisfied by $t=\kappa \tau$.
  The equation of motion for ${X}_a$ is 
\beq
-\partial^2_\tau X_a + \partial_\sigma^2 X_a - \Lambda X_a =0
\eeq
Motivated by the above remark we consider the following generalization of the NR ansatz in~\cite{ART}:
\beq
 X_a = x_a(\xi)\ e^{i\omega_a\tau}\ , \ \ \ \ \ \ \ \ 
 \ \ \   \ \ \xi\equiv \alpha \sigma + \beta \tau \ , 
 \eeq
where $x_a=r_a e^{i \mu_a}$ are in general complex and 
the  periodicity in $\sigma$ translates into the condition 
 \beq 
  x_a(\xi+2\pi\alpha)=x_a(\xi)\ . 
\eeq
Variations of this ansatz  describe also 
the spinning rigid strings \ci{AFRT}  and 
pulsating \ci{minahan} strings \ci{ART,KT}.\footnote{More generally, one may consider 
the ansatz $X_a = x_a(\xi)\ e^{i\omega_a\tau + i m_a \sigma}$. Then pulsating string case 
corresponds to $\alpha=0$, i.e. $x_a(\xi) \to x_a(\tau)$. For non-zero $\alpha$ 
the additional windings $m_a$  can be set to zero as they   can be absorbed into the 
phase of $x_a$.}

 The conformal constraints read
\beqa
\sum_a \left[ |\partial_\tau X_a|^2 + |\partial_\sigma X_a|^2\right ] = \kappa^2  \ , \ \ \ \ \ \ \ 
\sum_a\left[ \partial_\tau X_a \partial_\sigma \bar{X}_a + \partial_\tau \bar{X}_a \partial_\sigma X_a \right]= 0\ . 
\eeqa
We have  
\beqa 
\partial_\tau X_a = (\beta x'_a + i\omega_a x_a) e^{i\omega_a \tau} \ ,  \ \ \ \ \ \ \ \
\partial_\sigma X_a = \alpha x'_a e^{i\omega_a \tau}\ , 
\eeqa
 where primes denote derivatives with respect to $\xi$.
The equations of motion become
\beq
(\alpha^2-\beta^2) x''_a - 2 i \beta \omega_a x'_a +\omega_a^2 x_a - \Lambda x_a=0\ , 
\eeq
which follow from the following Lagrangian for $x_a$:
\beq
 \cL = \sum_a\left[ (\alpha^2-\beta^2) x'_a \bar{x}'_a + i\beta\omega_a (x'_a\bar{x}_a - \bar{x}'_a x_a) - \omega_a^2 x_a \bar{x}_a \right] 
 + \Lambda (\sum_a x_a \bar{x}_a -1)
\label{xLag}
\eeq
 Except for the term proportional to $\beta$, this Lagrangian is that of the 
  Neumann system. It describes the motion of a particle 
on a sphere under a quadratic potential and is integrable~\cite{NR, Moser}. 
The term proportional to $\beta$ can be described as a magnetic 
field and, as we shall see below, does not modify the radial (NR) equations.  
 Pictorially, a particle would like to oscillate as in the usual NR system but the 
magnetic field bends the trajectory giving rise to arcs. 
Since the form of the trajectory of this fictitious particle represents  the
 shape of the string, those are
the arcs between the spikes in the spiky string, and, in particular,  
the single arc  of the giant magnon.

The Hamiltonian corresponding to (\ref{xLag}) is (assuming $\sum_a x_a \bar{x}_a=1$)
\beq
H = \sum_a\left[(\alpha^2-\beta^2) x'_a \bar{x}'_a + \omega_a^2 x_a \bar{x}_a \right] \ . 
\eeq
Defining (no sum over $a$)\footnote{Notice that in this paper we write all summations explicitely.} 
\beq
\Xi_a = i (x'_a \bar{x}_a - \bar{x}'_a x_a)\ , 
\eeq 
we can rewrite the constraints as:
\beqa
 (\alpha^2-\beta^2) \sum_a x'_a \bar{x}'_a + \sum_a \omega_a^2 x_a \bar{x}_a  &=& \kappa^2\ ,  \\
 \frac{\alpha^2-\beta^2}{2\beta} \sum_a \omega_a\, \Xi_a +\sum_a \omega_a^2 x_a \bar{x}_a &=& \kappa^2\ . 
\eeqa
The first one is conserved since it is related to 
 the Hamiltonian. The second one  is conserved if we use the equations of motion, 
 implying, in particular, that 
\beq
 (\alpha^2 -\beta^2)\, \Xi'_a = -2\beta \omega_a (x_a\bar{x}_a)'\ . 
\eeq
This means that we have just to fix conserved quantities to satisfy the constraints.

 Let us now  use the following  ``polar'' parameterization of $x_a$ 
\beq
 x_a(\xi) = \rr_a(\xi) \ e^{i\mu_a(\xi)}\ , 
\eeq
where $\rr_a$ are real.  Then
\beqa
|x'_a|^2 &= &\rr'_a{}^2 + \rr_a^2 \mu'_a{}^2 \ , \\
\Xi_a &=& -2\rr_a^2 \mu'_a\ . 
\eeqa
The Lagrangian becomes:
\beq
\cL = \sum_a \bigg[ (\alpha^2-\beta^2)  \rr'_a{}^2 + (\alpha^2-\beta^2) 
\rr_a^2 \bigg(\mu'_a - \frac{\beta\omega_a}{\alpha^2-\beta^2}\bigg)^2
 - \frac{\alpha^2}{\alpha^2-\beta^2} \omega_a^2 \rr_a^2 \bigg] + \Lambda\big( \sum_a\rr_a^2 -1 \big)
\eeq
The equations of motion for $\mu_a$ are easily integrated,  giving:
\beq
\mu'_a = \frac{1}{\alpha^2-\beta^2} \left[\frac{C_a}{\rr_a^2} + \beta \omega_a\right]\ , 
\eeq
where $C_a$ are constants of motion. 
Using this in the equations of motion for $\rr_a$ 
we get
\beq
(\a^2-\beta^2) r_a''- {C_a^2\over (\a^2-\beta^2)}\, {1\over r_a^3}
+{\a^2\over (\a^2-\beta^2)}\, \omega_a^2r_a  - \Lambda r_a =0\ ,
\label{raaa}
\eeq
 which can be derived from  
the Lagrangian:
\beq
 \cL =\sum_a\bigg[ (\alpha^2-\beta^2) \rr'_a{}^2 - \frac{1}{\alpha^2-\beta^2}\frac{C_a^2}{\rr_a^2} 
        - \frac{\alpha^2}{\alpha^2-\beta^2}  \omega_a^2\rr_a^2\bigg]
         + \Lambda(\sum_a\rr_a^2-1)\ , 
\label{Lagrho}
\eeq
with the corresponding Hamiltonian being 
\beq
 H =  \sum_a \bigg[  (\alpha^2-\beta^2)\rr'_a{}^2 + 
 \frac{1}{\alpha^2-\beta^2}   \frac{C_a^2}{\rr_a^2} 
        + \frac{\alpha^2}{\alpha^2-\beta^2}  \omega_a^2\rr_a^2 \bigg]\ . 
\eeq
The constraints are satisfied if 
\beqa
 \sum_a\,\omega_a C_a + \beta \kappa^2 = 0 \ , \ \ \ \ \ \ \ \ \ 
 H =  \frac{\alpha^2+\beta^2}{\alpha^2-\beta^2} \kappa^2 \ . 
\eeqa
 The periodicity conditions read:
\beqa
 \rr_a(\xi+2\pi\alpha) = \rr_a(\xi) \ , \ \ \ \ \ \ \ \ \ 
   \mu_a( \xi+2\pi\alpha)= \mu_a(\xi) + 2 \pi n_a \ , \eeqa
 where $n_a$  are integer winding numers;
  the second condition implies 
  \beqa
 \frac{C_a}{2\pi} \int_0^{2\pi\alpha} \frac{d\xi}{\rr_a^2} =
  (\alpha^2 -\beta^2) n_a - \alpha\beta \omega_a \ .
\eeqa
The Lagrangian (\ref{Lagrho}) describes the standard NR integrable system.
Thus the general solution for our ansatz can be constructed in terms of the
usual solutions of the NR system. There are five independent integrals of
motion which reduce the equations to a system of first-order equations that
can be directly integrated \ci{AFRT}. In the next subsection, we shall 
present a direct
derivation of these integrals of motion for our particular case.

\subsection{Conserved quantities}

Let us start with  the Lagrangian (\ref{xLag}) and define the momenta as:
\beq
p_a = \frac{\partial \cL}{\partial\bar{x}_a} = (\alpha^2-\beta^2) x'_a 
- i\beta\omega_ax_a \ . 
\eeq
Then 
\beq
p'_a = i\beta\omega_ax'_a -\omega^2_a x_a -\Lambda x_a\ , 
\eeq
which implies
\beq
(\bar{x}_b p_a -x_a \bar{p}_b)' = \frac{i\beta}{\alpha^2-\beta^2} (\omega_a-\omega_b) (\bar{x}_bp_a-x_a\bar{p}_b)
           +\frac{\alpha^2}{\alpha^2-\beta^2} (\omega_b^2-\omega_a^2)x_a\bar{x}_b
\eeq
From here we obtain
\beq
\partial_\xi \sum_{b\neq a} \frac{1}{\omega_b^2-\omega_a^2} | \bar{x}_b p_a -x_a \bar{p}_b |^2 = \alpha^2 (x_a\bar{x}_a)'
\eeq
 which implies that the quantities
\beq
F_a = \alpha^2 x_a\bar{x}_a + \sum_{b\neq a} \frac{| \bar{x}_b p_a -x_a \bar{p}_b |^2}{\omega_a^2-\omega_b^2}  
\eeq
are conserved. They are not all independent since $\sum_a F_a = \alpha^2$.
Expressed in  terms of the radii $\rr_a$ they read:
\beq
 F_a = \alpha^2 \rr_a^2 + (\alpha^2-\beta^2)^2 \sum_{b\neq a} \frac{(\rr_b\rr_a'-\rr_a\rr_b')^2}{\omega_a^2-\omega_b^2} 
      +\sum_{b\neq a} \frac{1}{\omega_a^2-\omega_b^2}\left(\frac{C_a\rr_b}{\rr_a}+\frac{C_b\rr_a}{\rr_b}\right)^2
\label{Fr}
\eeq
 Notice, in particular, from the last term, that if a
  certain solution reaches a point where some $\rr_a=0$ then we should have the 
corresponding $C_a=0$. 
Later we are going to find a solution which reaches the 
point $(\rr_1,\rr_2,\rr_3)=(1,0,0)$,  where $\rr'_a=0$.
It then  follows immediately that $C_{2,3}=0$ and $F_1=\alpha^2$, $F_{2,3}=0$.

 We now have three conserved quantities $C_a$ and another two among the $F_a$ since only two $F_a$  are independent. It is
important to write the Hamiltonian in terms of the conserved quantities. We get after some simple algebra:
\beq
H = \frac{1}{\alpha^2-\beta^2} \bigg[
 \sum_a\bigg(  \omega_a^2 F_a + 2\beta \omega_a C_a + 2 C_a^2\bigg)
  -\bigg(\sum_a C_a\bigg)^2\bigg] \ . 
\eeq
The conformal constraints imply a closely related expression
\beq
 (\alpha^2+\beta^2) \kappa^2 = \sum_a \big(  \omega_a^2 F_a +  C_a^2\big) 
  - \sum_{a\neq b} C_a C_b\ . 
\label{kappaFrel}
\eeq
Note that the  characteristic frequencies of the motion are the
 derivatives of the Hamiltonian with respect to the conserved momenta. Therefore, 
we can directly compute them from the above  expression.
%
%
%

\subsection{Angular momenta}

 The original lagrangian (\ref{orlag}) is invariant under $SO(6)$ rotations. We  can define the conjugate momenta to $\bar{X}_a$ 
 as  $\Pi_a=\dot{X}_a$ and then for  the (complex) angular momentum 
 components we get  (and similar expressions for their complex conjugate components)
\beqa
 J_{a\bar{b}} &=& T \int d\sigma \left( X_a \Pi_{\bar{b}} - X_{\bar{b}} \Pi_a \right) \ , \\
 J_{ab} &=& T \int d\sigma \left( X_a \Pi_b - X_b \Pi_a \right)\ , 
\eeqa
where  $T= { \sqrt \lambda \over 2 \pi} $ 
is string tension which  appears in front of the string action.
 Using our ansatz for $X_a$  we get
\beqa
 J_{a\bar{b}} &=& Te^{i(\omega_a-\omega_b)\tau}\ 
             \int \frac{d\xi}{\alpha} \left[\beta(x_a\bar{x}'_b-\bar{x}_bx'_a)-i(\omega_a+\omega_b)x_a\bar{x}_b \right]\ ,  \\
 J_{ab} &=& T e^{i(\omega_a+\omega_b)\tau}\ 
             \int \frac{d\xi}{\alpha} \left[\beta(x_a{x}'_b-{x}_bx'_a)-i(\omega_b-\omega_a)x_ax_b \right]\ . 
\eeqa
 These must be time-independent quantities. However, 
  the time dependence  appears  not to cancel except for $J_{a\bar{a}}$
  (assuming all frequences  $\omega_a$ are different).
 This means  that  the coefficients 
 multiplying the time-dependent exponential factors should actually vanish. 
 As a result, only the diagonal (Cartan)
 components of the angular momentum tensor may be non-zero for the
 solutions described by the  NR ansatz (the same  argument was given in \cite{AFRT}) 
\beq
 J_a \equiv  J_{a\bar{a}} = T \int  d\xi \bigg( 
 \frac{\beta}{\alpha} \frac{C_a}{\alpha^2-\beta^2}  + \frac{\alpha  \omega_a }{\alpha^2-\beta^2} \rr_a^2\bigg)
 \ .
\eeq
Here we have used that 
 $x_a = \rr_a e^{i\mu_a}$ as well as  the equations of motion for $\mu_a$. 
 If we further notice that the energy of the string is given by
\beq
E = T {\kappa\over \alpha}  \int d\xi
\eeq
we obtain a relation
\beq
\frac{\alpha^2+\beta\sum_a\frac{C_a}{\omega_a}}{\alpha^2-\beta^2}
 \frac{ E}{ \kappa} = \sum_a \frac{J_a}{\omega_a}\ . 
\label{kappaJrel}
\eeq 
 Finally, let us  comment  on  the limits of the integrals over  $\xi$.
 For standard  closed strings with $0\le \sigma \le 2\pi$ 
 we have $0\le \xi\le 2\pi\alpha$. However, for strings with 
 infinite energy and momenta with $E-J$ fixed as in \cite{HM}
 one has $\kappa \to \infty$ and then it is natural to rescale 
 $\xi$ so that it takes values on an infinite line; equivalently,
 in this case  we may keep $\kappa$ finite (or set  $\kappa=1$)
  while assuming that 
 $-\infty\le \xi \le \infty$.


\section{A solution with two angular momenta}

 A giant magnon solution with one infinite and one finite 
  angular momentum  on $S^3$ was found in \cite{Dorey2,AFZ,MTT}. 
 Here we shall reproduce it using   our NR  ansatz. 
We shall use the  expressions of the previous section (with $a=1,2$) 
but set $$\alpha=1$$
 to  simplify  the notation. We have the constraints 
\beq
\omega_1 C_1+\omega_2 C_2 +\beta \kappa^2 =0, \ \ \ \ \ \ \ \ \ \ \ \ H
 = \frac{1+\beta^2}{1-\beta^2} \kappa^2 \ . 
\eeq
Using that $H$ is conserved and that here $\rr_1^2+\rr_2^2 =1$ we  immediately 
find the solution. We get 
\beqa
H 
       = (1-\beta^2) \frac{\rr_1'{}^2 }{1-\rr_1^2} + \frac{1}{1-\beta^2} \left(\frac{C_1^2}{\rr_1^2}+\frac{C_2^2}{1-\rr_1^2}\right)
           +\frac{\omega_1^2-\omega_2^2}{1-\beta^2} \rr_1^2 + \frac{\omega_2^2}{1-\beta^2} 
\eeqa
From here (and the relation  $H = \frac{1+\beta^2}{1-\beta^2} \kappa^2$) 
we obtain 
\beqa
&&(1-\beta^2)^2 \rr_1'{}^2 = \nonumber\\
&&\ =  \frac{1}{\rr_1^2}\left[((1+\beta^2)\kappa^2-\omega_2^2)\rr_1^2(1-\rr_1^2)-C_1^2+(C_1^2-C_2^2)\rr_1^2 -(\omega_1^2-\omega_2^2)\rr_1^4(1-\rr_1^2)  \right] \nonumber 
\eeqa
The right hand side has three zeros which correspond to turning points where $\rr_1'=0$. We want one of them to be $\rr_1=1$ so that the string
extends to the equator. 
Replacing $\rr_1$ by 1 in the right hand side we get  zero only if $C_2=0$, so this  determines this constant of motion. The
equation then simplifies to:
\beq
(1-\beta^2)^2 \rr_1'{}^2 = 
        \frac{1-\rr_1^2}{\rr_1^2}\left[((1+\beta^2)\kappa^2-\omega_2^2)\rr_1^2-C_1^2 -(\omega_1^2-\omega_2^2)\rr_1^4  \right]
\eeq
However,  we
still get two zeros. It turns out that one needs $\rr_1=1$ to be a double zero. 
Replacing $\rr_1$ in the right hand side we
get $(1+\beta^2) \kappa^2 = \omega_1^2 + C_1^2$
and using  that $C_2=0$ we get $\beta=-\frac{\omega_1C_1}{\kappa^2}$ 
which then implies $
 \kappa^4+\omega_1^2C_1^2=\omega_1^2\kappa^2+C_1^2\kappa^2
$. 
Solving for $\kappa$ we get\footnote{We assume that 
the sign choices are such that the energy and the spins are positive.}
 $\kappa=\omega_1$ 
or $\kappa=C_1$. 
We will see later that the first choice $\kappa=\omega_1$ is the 
required  one to get a giant magnon.
The equation for $\rr_1$ is then  further simplified to:
\beq
(1-\beta^2)^2 \rr_1'{}^2 =  \frac{(1-\rr_1^2)^2}{\rr_1^2} (\omega_1^2-\omega_2^2) (\rr_1^2-\bar{\rr}_1^2)\ , \ \ 
\eeq
where 
\beq
\bar{\rr}_1=\frac{C_1}{\sqrt{\omega_1^2-\omega_2^2}}
\eeq
is the other turning point that determines the extension of the string. 
Equivalently, this equation may be written as
\beq
u' = \frac{2}{1-\beta^2} (1-u) \sqrt{u-\bar{u}}\sqrt{\omega_1^2-\omega_2^2}\ , 
\ \ \ \ \ \ \ \ \ \ \ 
u\equiv  \rr_1^2 \ , \ \ \ \  \bar u\equiv \bar  \rr_1^2 \ .
\eeq
The conserved charges  are:
\beqa
 E &=& \kappa T \int d\xi \\
 J_1 &=& \frac{\beta C_1}{1-\beta^2} T \int d\xi + \frac{\omega_1}{1-\beta^2} T \int ud\xi \\
 J_2 &=& \frac{\omega_2}{1-\beta^2}T \int (1-u) d\xi \ . 
\eeqa
The angular extension of the string is
\beq
\dmu = \int \mu'_1 \, d\xi=\frac{C_1}{1-\beta^2} \int\frac{d\xi}{u} + \frac{\beta\omega_1}{1-\beta^2}\int d\xi
\eeq
 A simple computation using that $\beta\omega_1=-C_1\omega_1^2/\kappa^2=-C_1$ gives a finite result. This justifies
the choice $\kappa=\omega_1$ in the previous equation for $\kappa$. The result is
\beq
 \dmu = \frac{2\,C_1}{\sqrt{\bar{u}}\sqrt{\omega_1^2-\omega_2^2}}\mbox{arccos}\sqrt{\bar{u}} = 2\,\mbox{arccos}\sqrt{\bar{u}}\ . 
\eeq
The angular momenta can be computed using  that
\beq
 \int (1-u) d\xi = 2 \int_{\bar{u}}^1 \frac{1-u}{u'}du = 2 \frac{1-\beta^2}{\sqrt{\omega_1^2-\omega_2^2}}\sqrt{1-\bar{u}}
\eeq
The factor of two is because the integral between $\bar{u}$ and $1$ is only half of the string. We obtain:
\beqa
J_1 &=& \frac{\beta C_1 + \omega_1 }{1-\beta^2} \frac{E}{\kappa}
   -\frac{2\,\omega_1\,T}{\sqrt{\omega_1^2-\omega_2^2}} \sqrt{1-\bar{u}}
    = E -\frac{2\,\omega_1\,T}{\sqrt{\omega_1^2-\omega_2^2}} \sqrt{1-\bar{u}}\ ,  \\
J_2 &=& \frac{2\, \omega_2\,T}{\sqrt{\omega_1^2-\omega_2^2}}\sqrt{1-\bar{u}}\ . 
\eeqa
where we used that $\omega_1=\kappa$ and $\beta\omega_1=-C_1$. 
We can write the charges  in terms of the angle $\dmu$
 and an auxiliary  angle $\gamma$ defined by  $\omega_2=\kappa\sin \gamma$.
Observing that 
\beq
\sqrt{1-\bar{u}} = \sin \frac{\dmu}{2}, \ \ \ \sqrt{\omega_1^2-\omega_2^2}=\kappa \cos\gamma\ , 
\eeq
we get 
\beqa
 \Delta \equiv  E-J_1 = 2T\,\frac{\sin\frac{\dmu}{2}}{\cos\gamma} \ , \ \ \ \ \ \ \ \ \ \ \ 
 J_2 = 2T\,\sin\frac{\dmu}{2} \tan\gamma \label{J2}\ . 
\eeqa
 Then 
\beq
\Delta^2 = J_2^2 + 4T^2\sin^2 \frac{\dmu}{2}\ . 
\label{Delta}
\eeq
 Finally, using 
  that the string tension is $T=\frac{\sqrt{\thc}}{2\pi}$  we arrive at:
\beq
\Delta = \sqrt{J_2^2 + \frac{\thc}{\pi^2} \sin^2 \frac{\dmu}{2}} \ ,  
\label{twoJJ}
\eeq 
which is the same energy relation as in \cite{Dorey2} after we identify 
$\dmu$ with the giant magnon momentum $p$ as in \cite{HM}.
 Notice  also  that using $C_{2,3}=0$ (and $J_3=0$) we get from (\ref{kappaJrel}):
\beq
 E = J_1 + \frac{\kappa}{\omega_2} J_2 \ ,  \ \ \ \ \ \  {\rm i.e.} \ \ \ \ \ 
  \ \ \Delta = E-J_1 = \frac{J_2}{\sin\gamma}\ , 
\eeq
which is consistent with  (\ref{J2}). 

 It is interesting  to compute  the   NR integrals of motion 
  $F_a$ correspondng  to this giant magnon solution. Using eq.(\ref{Fr}) at the point $\rr_1=1$, $\rr_2=\rr_3=0$,
$\rr'_a=0$,  we get simply:
\beq
 F_1=1, \ \ \ F_2=F_3=0 \ . 
\eeq 
 A  simple check is that eq. (\ref{kappaFrel}) reduces to the relation 
 $(1+\beta^2) \kappa^2 = \omega_1^2 + C_1^2$ which we found above.

\section{A solution with three angular momenta}
Here we shall find a new  giant magnon solution with two 
extra angular momenta.

\subsection{Form of the  solution}

 To get a solution with three non-zero
 angular momenta we put all $\omega_a\neq0$   and change from the three constrained 
 radial  variables 
$\rr_a$ to two unconstrained ones $\lapm$  (as is standard when solving the NR system \cite{Moser,AFRT}):
\beq
\sum_{a=1}^3 \frac{\rr_a^2}{\la-\omega_a^2} = \frac{(\la-\lap)(\la-\lam)}{\prod_{a=1}^3 (\la-\omega_a^2)}\ . 
\label{lambdadef}
\eeq
 $\lapm$ are the roots of the quadratic equation 
 obtained by taking common denominator on the left hand side and equating the
numerator to zero. The two roots are such that $\omega_3^2<\lam<\omega_2^2<\lap<\omega_1^2$. They satisfy:
\beq
 \lap+\lam = \sum_a \omega_a^2 - \sum_a \omega_a^2 \rr_a^2 , \ \ \ \ \ \lap\lam = \prod_a\omega_a^2\ \times   \sum_b \frac{\rr_b^2}{\omega_b^2}\ , 
\label{lambdaid}
\eeq
as follows from equating the left and right hand side of eq.(\ref{lambdadef}). We can invert this  transformation  to get
\beq
 \rr_a^2 = \frac{(\lap-\omega_a^2)(\lam-\omega_a^2)}{\prod_{b\neq a} (\omega_a^2-\omega_b^2)}
\eeq
 A straightforward computation then gives the Lagrangian in terms of $\lapm$ (again we set $\alpha=1$):
\beqa
 \cL &=&  { 1 \over 4}(1-\beta^2) (\lap-\lam) \left( \frac{\lam'{}^2}{\prod_a(\lam-\omega_a^2)} 
  - \frac{\lap'{}^2}{\prod_a(\lap-\omega_a^2)} \right) \nonumber \\ 
&& - \frac{1}{(1-\beta^2)}\frac{1}{(\lap-\lam)}  \left( \sum_a \prod_{b\neq a} (\omega_a^2-\omega_b^2) 
 \left[\frac{C_a^2}{\lam-\omega_a^2}-\frac{C_a^2}{\lap-\omega_a^2}\right]\right)\nonumber  \\
&&  -\frac{1}{1-\beta^2} \left( \sum_a\omega_a^2-(\lap+\lam)\right)
\eeqa
 and the Hamiltonian
\beqa
 H _{\la} &=& \frac{1}{(1-\beta^2)(\lap-\lam)} \left\{\tilde{H}(p_-,\lam) - \tilde{H}(p_+,\lap)\right\}\ ,  \\
 \tilde{H}(p,\la) &=& \prod_a(\la-\omega_a^2)\ p^2 
           + \sum_a C_a^2 \frac{\prod_{b\neq a} (\omega_a^2-\omega_b^2)}{\la-\omega_a^2} + \sum_a\omega_a^2\ \la - \la^2 \ . 
\eeqa
 One way to study this system is to use the Hamilton-Jacobi method which requires finding a function $\cW(\lap,\lam)$ such that
\beq
 H_\la\left(p_\pm=\frac{\partial \cW}{\partial \lapm},\lapm\right) = E\ . 
\eeq
 If a solution of the form $\cW = W_+(\lap) + W_-(\lam)$ exists we say that the variables separate and the system is integrable in these coordinates.  
Trying such a solution in our case 
 we obtain that, in fact, $W_\pm$ are the same function obtained from integrating
 the equation 
\beq
 \left(\frac{\partial W}{\partial \la}\right)^2  = \frac{\left\{ V  - \sum_a \prod_{b\neq a} (\omega_a^2-\omega_b^2)
 \frac{C_a^2}{\la-\omega_a^2} 
   + \left[\kappa^2 (1+\beta^2) - \sum_a\omega_a^2\right] \la - \la^2\right\} }{\prod_a(\la-\omega_a^2)}\nonumber
\eeq
 where $V$ is a constant of motion and we used the relation $E=\frac{1+\beta^2}{1-\beta^2}\kappa^2$. 
The solution of the Hamilton-Jacobi equation is then 
\beqa  \cW(\lapm,V,E) = W(\lap,V,E)+W(\lam,V,E)\ .  \eeqa 
 The equations of motion reduce to
\beqa
 \frac{\partial W(\lap,V,E)}{\partial V} +  \frac{\partial
  W(\lam,V,E)}{\partial V} &=& U\ ,  \\
 \frac{\partial W(\lap,V,E)}{\partial E} +  \frac{\partial
  W(\lam,V,E)}{\partial E} &=& \xi \ . 
\eeqa
where $U$ is a new constant. The first equation determines $\lap$ as a function of $\lam$, 
 and the second equation determines how  both of them depend on 
 the `time'  variable $\xi$.
Computing the derivatives of $W$ we find 
\beqa
  \int^{\lap} \frac{d\la}{\sqrt{P_5(\la)}} +   \int^{\lam} \frac{d\la}{\sqrt{P_5(\la)}} &=& 2U \label{sol1a}\ ,  \\
  \int^{\lap} \frac{\la\, d\la}{\sqrt{P_5(\la)}} +   \int^{\lam} \frac{\la\, d\la}{\sqrt{P_5(\la)}} &=& -\frac{2\xi}{1-\beta^2} \ , 
\label{sol1}
\eeqa
where we 
defined the quintic polynomial $P_5(\la)$ as:
\beq
P_5(\la) = \prod_a(\la-\omega_a^2)
        \bigg\{ V  - \sum_a \prod_{b\neq a} (\omega_a^2-\omega_b^2) \frac{C_a^2}{\la-\omega_a^2} 
   + \bigg[\kappa^2 (1+\beta^2) - \sum_a\omega_a^2\bigg] \la - \la^2 \bigg\}
\eeq
 Although one could use (\ref{sol1a}),(\ref{sol1}) to find the shape of the 
 generic string solution, here we are interested in particular 
 solutions describing strings with one infinite momentum (or ``infinitely long'' strings). 
Such solutions arise when $\la_\pm$ can reach its extremal values $\omega_{2,3}^2$. For this to happen we choose $V$ and $E$ (or $\kappa$) such that
$P_5(\la)$ has a double zero at $\la=\omega_2^2$ and a double zero at
 $\la=\omega_3^2$. For this we  need to choose
\beq
C_2=0, \ \ \ C_3=0, \ \ \ \kappa^2(1+\beta^2) = \omega_1^2+C_1^2, \ \ \ V = -\omega_2^2\omega_3^2 - C_1^2(\omega_1^2-\omega_2^2-\omega_3^2)\ . 
\eeq
As in the 2-spin case, if we use the conformal
 constraints this implies 
\beq
\omega_1=\kappa, \ \ \ \ \beta=-\frac{C_1}{\omega_1}. 
\eeq
 The equations to solve then reduce to
\beqa
\int^{\lap}_{\lab} \frac{d\la}{(\la-\omega_2^2)(\la-\omega_3^2)\sqrt{\lab-\la}} +  
 \int^{\lam}_{\labm} \frac{d\la}{(\la-\omega_2^2)(\la-\omega_3^2)\sqrt{\lab-\la}} &=& 0 \\
\int^{\lap}_{\lab} \frac{\la\, d\la}{(\la-\omega_2^2)(\la-\omega_3^2)\sqrt{\lab-\la}} +  
 \int^{\lam}_{\labm} \frac{\la\, d\la}{(\la-\omega_2^2)(\la-\omega_3^2)\sqrt{\lab-\la}} &=& -\frac{2\xi}{1-\beta^2} \ , 
\eeqa
which can be integrated  by elementary methods. 
Here $\lab = \sqrt{\omega_1^2-C_1^2}$ \  ($\omega_2^2<\lab<\omega_1^2$)\footnote{We assume that 
$C_1^2<w_1^2-w_2^2$ since  otherwise there is no solution.} is the maximum value of $\lap$ and we assume that at such point $\lam$ has 
an arbitrary value $\labm$\  ($\omega_2^2<\labm<\omega_3^2$). Changing $\labm$ changes the integral by a constant and that allowed us to
absorb $U$ in the definition of $\labm$.

The above equations can be simplified to 
\beqa
 \int^{\lap}_{\lab}  \frac{d\la}{(\la-\omega_3^2)\sqrt{\lab-\la}} +  
 \int^{\lam}_{\labm} \frac{d\la}{(\la-\omega_3^2)\sqrt{\lab-\la}} &=& -\frac{2\xi}{1-\beta^2}\ ,  \label{om3int}\\
 \int^{\lap}_{\lab}  \frac{d\la}{(\la-\omega_2^2)\sqrt{\lab-\la}} +  
 \int^{\lam}_{\labm} \frac{d\la}{(\la-\omega_2^2)\sqrt{\lab-\la}} &=& -\frac{2\xi}{1-\beta^2} \ . \label{om2int}
\eeqa 
 We  find then 
\beqa
  \int^{\lap}_{\lab}  \frac{d\la}{(\la-\omega_2^2)\sqrt{\lab-\la}} &=& \frac{2}{\sqrt{\lab-\omega_2^2}} 
       \mbox{arctanh}\frac{\sqrt{\lab-\lap}}{\sqrt{\lab-\omega_2^2}} \\
  \int^{\lam}_{\labm} \frac{d\la}{(\la-\omega_2^2)\sqrt{\lab-\la}} &=& \frac{2}{\sqrt{\lab-\omega_2^2}} \bigg[
       \mbox{arctanh}\frac{\sqrt{\lab-\omega_2^2}}{\sqrt{\lab-\lam}}
              -
       \mbox{arctanh}\frac{\sqrt{\lab-\omega_2^2}}{\sqrt{\lab-\labm}}\bigg] ,  \nonumber
\eeqa
 which are slightly different because $\lap>\omega_2^2$ and $\lam<\omega_2^2$ (also, 
  the limits of integration are different).
In a similar way we can do the integrals in eq.(\ref{om3int}) taking into account that $\la_\pm>\omega_3^2$.
 Using  these results we find the following  algebraic equations
\beqa
 \frac{s_+s_-+s_2^2}{s_++s_-}&=& s_2A_2(\xi)\\
 \frac{s_+s_-+s_3^2}{s_++s_-}&=& s_3A_3(\xi)
\eeqa
where we defined ($s_1$ is introduced here  for later use)
\beq
 s_1 = \sqrt{w_1^2-\lab}, \ \ \ \  s_{2,3} = \sqrt{\lab-\omega_{2,3}^2}, \ \ \ \ s_{\pm} = \sqrt{\lab-\lapm}, 
\eeq
\beq
 A_2(\xi) = \tanh\left( - \frac{s_2 \xi}{1-\beta^2} +B_2 \right), \ \ \ \ \ \ \ \ 
  A_3(\xi) = \coth\left( - \frac{s_3 \xi}{1-\beta^2} +B_3 \right)\ . \label{aa}
\eeq
 Here we defined:
\beq
 \tanh B_2 = \frac{s_2}{\sqrt{\lab-\labm}}, \ \ \ \ \ \ \ \ \ 
   \tanh B_3 = \frac{\sqrt{\lab-\labm}}{s_3}\ , 
\eeq
and $\xi$ is assumed to extend from $-\infty$ to $+\infty$. 
 To go back to the variables $\rr_a$ we note  that 
\beq
 \rr_a^2 = \frac{(\lap-\omega_a^2)(\lam-\omega_a^2)}{\prod_{b\neq a}(\omega_a^2-\omega_b^2)} 
          = \frac{(s_a^2-s_+^2)(s_a^2-s_-^2)}{\prod_{b\neq a}(\omega_a^2-\omega_b^2)}
          = \frac{(s_a^2+s_+s_-)^2-s_a^2(s_++s_-)^2}{\prod_{b\neq a}(\omega_a^2-\omega_b^2)}\ . 
\eeq
 Using that
\beqa
 s_++s_- = -\frac{\omega_2^2-\omega_3^2}{s_3A_3(\xi)-s_2A_2(\xi)} \ , \ \ \ \ \ \ \ \ \ 
 s_+s_-  = s_2s_3\,\frac{s_3A_2(\xi)-s_2A_3(\xi)}{s_3A_3(\xi)-s_2A_2(\xi)}   
\label{spsm}
\eeqa
 this results in 
\beqa
  \rr_1^2 &=& \frac{\left[(\omega_1^2-\omega_2^2)s_3A_3(\xi)-(\omega_1^2-\omega_3^2)s_2A_2(\xi)\right]^2+s_1^2(\omega_2^2-\omega_3^2)^2}
                    {(\omega_1^2-\omega_2^2)(\omega_1^2-\omega_3^2)(s_3A_3(\xi)-s_2A_2(\xi))^2} \\ \nonumber\\
  \rr_2^2 &=& \frac{(\omega_2^2-\omega_3^2)}{(\omega_1^2-\omega_2^2)} s_2^2 \frac{1-A_2^2(\xi)}{(s_3A_3(\xi)-s_2A_2(\xi))^2} \\ \nonumber\\
  \rr_3^2 &=& \frac{(\omega_2^2-\omega_3^2)}{(\omega_1^2-\omega_3^2)} s_3^2
   \frac{ A_3^2(\xi) - 1  }{(s_3A_3(\xi)-s_2A_2(\xi))^2} \ . 
\eeqa
Together with (\ref{aa}) this 
 gives explicitly $\rr_{a}$ as simple 
  functions of  $\xi$. 
  It is easy to check that 
$\sum_a\rr_a^2=1$ and $\rr^2_a \ge 0$ \ $(a=1,2,3)$. 

 One  can also check directly 
that the equations of motion for $\rr_a$ following from the Lagrangian (\ref{Lagrho}) are satisfied.\footnote{The coordinates 
$\lapm$ can at this point be ignored and one can work 
directly with the solution $\rr_a(\xi)$ that we obtained. As we have shown, 
$\lapm$ are, however,  important to derive the solution.}

\subsection{Energy and momenta}

 Since here $C_{2,3}=0$, the angular momenta $J_{2,3}$ can be computed as
\beq
 J_{a} = \frac{T}{1-\beta^2} \int_{-\infty}^{+\infty} \omega_a \rr^2_a(\xi) \ d\xi, \ \ \ \ \ \ \ a=2,3
\eeq
 Using the explicit expresions for $\rr_a(\xi)$
   and  the integrals 
\beqa
 \int_{-\infty}^{+\infty} \frac{(1-\tanh^2(x))\ dx}{[\tanh(x)-c \coth(c x+b)]^2} &=& \frac{2}{c^2-1} \\
 \int_{-\infty}^{+\infty} \frac{(\coth^2(cx+b)-1)\ dx}{[\tanh(x)-c \coth(c x+b)]^2} &=& \frac{2}{c(c^2-1)} 
\eeqa
we obtain that:
\beq
\frac{1}{T} J_a = \frac{2\,\omega_a s_a}{\omega_1^2-\omega_a^2} = \frac{2\,\omega_a}{\omega_1^2-\omega_a^2}\sqrt{\lab-\omega_a^2}, \ \ \ \ \ \ \ \ \  a=2,3 
\eeq
 The remaining
  angular momentum $J_1$ follows from the formula (\ref{kappaJrel}) (remembering that $C_{2,3}=0$, $C_1=-\beta\omega_1$):
\beq
 \frac{E}{\kappa} = \sum_a\frac{J_a}{\omega_a} \ \ \ \Rightarrow \ \ \ \Delta = E- J_1 = \frac{\omega_1}{\omega_2} J_2 + \frac{\omega_1}{\omega_3} J_3
\eeq
 Notice that as in the  two-spin case 
 both $E=\kappa T\int_{-\infty}^{+\infty} d\xi$
  and $J_1$ diverge for this solution  but their 
difference $\Delta$ is finite.

Now let us  compute $\dmu$ that we associate with the momentum of the magnon
~\cite{HM}.\footnote{Note that  since $C_{2,3}=0$,  one finds that  $\mu'_{2,3}$ 
are constant and therefore $(\Delta\mu)_{2,3}=\int_{-\infty}^{\infty} \mu'_{2,3}$ are infinite.}
We get
\beq
 \dmu = \int_{-\infty}^{+\infty}\mu'_1 d\xi = \frac{C_1}{1-\beta^2} \int_{-\infty}^{+\infty} \frac{1-\rr_1^2}{\rr_1^2} d\xi
\eeq
where we used the equation for $\mu_1$  from  section 2  
 and the relation  $\beta=-\frac{C_1}{\omega_1}$. 
 This integral is convergent
   since $\rr_1$ approaches   1
exponentially fast as $\xi\rightarrow \pm\infty$.
 If we remember that
\beqa
\frac{2}{1-\beta^2} d\xi &=& \frac{\lap d\lap}{(\lap-\omega_2^2)(\lap-\omega_3^2)\sqrt{\lab-\lap}}+
          \frac{\lam d\lam}{(\lam-\omega_2^2)(\lam-\omega_3^2)\sqrt{\lab-\lam}}\\
 0    &=& \frac{d\lap}{(\lap-\omega_2^2)(\lap-\omega_3^2)\sqrt{\lab-\lap}}+
          \frac{d\lam}{(\lam-\omega_2^2)(\lam-\omega_3^2)\sqrt{\lab-\lam}}
\eeqa
we find that, in terms of the variables $\lapm$, 
\beqa
\frac{2}{C_1} d\mu_1 &=&  \left[-\frac{(\omega_1^2-\omega_2^2)(\omega_1^2-\omega_3^2)}{(\lap-\omega_1^2)(\lap-\omega_2^2)(\lap-\omega_3^2)}
           -\frac{\lap}{(\lap-\omega_2^2)(\lap-\omega_3^2)}\right]\frac{d\lap}{\sqrt{\lab-\lap}} \nonumber\\
   && + \left[-\frac{(\omega_1^2-\omega_2^2)(\omega_1^2-\omega_3^2)}{(\lam-\omega_1^2)
   (\lam-\omega_2^2)(\lam-\omega_3^2)}
           -\frac{\lam}{(\lam-\omega_2^2)(\lam-\omega_3^2)}\right]\frac{d\lam}{\sqrt{\lab-\lam}}  \nonumber\\
                             &=&  -\frac{d\lap}{(\lap-\omega_1^2)\sqrt{\lab-\lap}}
                                  -\frac{d\lam}{(\lam-\omega_1^2)\sqrt{\lab-\lam}}\ .
\eeqa
 Integrating over $\lapm$ we obtain:
\beq
 \mu_1 = -\arctan   \frac{\sqrt{\lab-\lap}}{\sqrt{\omega_1^2-\lab}}
         - \arctan  \frac{\sqrt{\lab-\lam}}{\sqrt{\omega_1^2-\lab}}
\eeq
 This can be written also as
\beq
 \tan \mu_1 = -\frac{s_1(s_2+s_3)}{s_1^2+s_+s_-}
\eeq
 which, through (\ref{spsm}) gives $\mu_1$ explicitly as a function of $\xi$. 
 Although this was derived for a piece of the string it can again be extended to all values $-\infty<\xi<\infty$.  In particular, since from (\ref{spsm})
we learn that $(s_+s_-)(\pm\infty) = s_2s_3$ and $(s_++s_-)(\pm\infty)=\pm \frac{\omega_2^2-\omega_3^2}{s_3-s_2}$, we find that
\beq
 \dmu = \mu_1(+\infty)-\mu_1(-\infty) 
 = 2\arctan \frac{s_1(s_2+s_3)}{s_1^2+s_2s_3}\ ,  
\eeq
 which can be written in the form:
\beq
\frac{\dmu}{2} = \arctan \frac{s_2}{s_1} + \arctan \frac{s_3}{s_1}\ .
\eeq
 Defining two angles $\phi_{2,3}$ by (below $a=2,3 $)
\beq
 \tan \phi_a =   \frac{s_a}{s_1}\ , \ \ \ \  \ \ \  0< \phi_a < \frac{\pi}{2} \ , 
 \  
\eeq
and another two $\gamma_{2,3}$ by
\beq
 \omega_{a} = \omega_1 \sin \gamma_{a} \ ,\ \ \ \   \  \ \ \ 0<\gamma_a<\frac{\pi}{2}\ , 
\eeq
we get 
\beq
 s_a = \sqrt{\omega_1^2-\omega_a^2}\, \sin\phi_a , \ \ \ 
\ \ \ \ \ \ \  J_a=2T\,\tan\gamma_a\sin\phi_a, \ \ 
\eeq
 Then 
\beq
\Delta = \frac{J_2}{\sin\gamma_2} + \frac{J_3}{\sin\gamma_3} = 2T \bigg( \,\frac{\sin\phi_2}
{\cos\gamma_2} + \frac{\sin\phi_3}{\cos\gamma_3}\bigg) \ .
\eeq
If we eliminate the variables $\gamma_a$ we obtain the final result
\beq
 \Delta = \sqrt{J_2^2 + \frac{\thc}{\pi^2}\sin^2\phi_2} +
  \sqrt{J_3^2 + \frac{\thc}{\pi^2}\sin^2\phi_3} \ , \ \ \ \ \ \ \ \ 
  \dmu = 2 ( \phi_2+\phi_3 ) \ , 
\label{eeefor}
\eeq
where we used that $T=\frac{\sqrt{\lambda}}{2\pi}$.
 The sum of $\phi_2$, $\phi_3$ is fixed but one might 
 wonder if they can otherwise be chosen arbitrarily. This is not 
 the case  if we keep  $J_{2,3}$ (or
$\omega_{2,3}$) fixed. Indeed, we have 
\beq
 \frac{1}{\cos^2\phi_a} = 1+ \tan\phi_a^2 = 1+ \frac{s_a^2}{s_1^2} =
 \cos^2\gamma_a\, \frac{\omega_1^2}{s_1^2}\ ,\qquad a=2,3\ , 
\eeq
and so 
\beq
 s_1 \sin\phi_2 = \omega_1 \cos\gamma_a\cos\phi_a\sin\phi_2 \ ,\qquad
 s_1 \sin\phi_3 = \omega_1 \cos\gamma_a\cos\phi_a\sin\phi_3 \ .
\eeq
If both $\phi_2 $ and $\phi_3$ are non-vanishing, this implies
the constraint
\beq
 \cos\gamma_2\cos\phi_2 = \cos\gamma_3 \cos\phi_3
\label{jjhh}
\eeq
 We can eliminate $\gamma_a$ in favor of $J_a$ obtaining the relation:
\beq
 \frac{\sin(2\phi_2)}{\sqrt{J_2^2 +{\lambda \over \pi^2} 
 \sin^2\phi_2}} =  \frac{\sin(2\phi_3)}{\sqrt{J_3^2 +   {\lambda \over \pi^2}  \sin^2\phi_3}}\ . 
\label{lkj}
\eeq
When either $\phi_2$ or $\phi_3$ vanishes, there is no
constraint.

Notice that the constraint (\ref{lkj}) can also be written as
\beq
 \frac{\partial \Delta_2}{\partial \phi_2} = \frac{\partial
   \Delta_3}{\partial \phi_3},\ \ \ \ \ \ \ \ \ \Delta_a \equiv  \sqrt{J_a^2 +
   { \lambda \over \pi^2} \sin^2\phi_a}, \ \ \ \ \ \ \ \ \ a=2,3\ .
\label{cond_f}
\eeq
Anticipating the result of the next  section,
 we are going to interpret this  solution as
 representing  two magnons with momenta $p_a = 2\phi_a$ and 
 energies $\Delta_a$. 
The classical configuration then 
describes two wave packets each 
 with group velocity $v_a = \half \frac{\partial{\Delta_a}}{\partial \phi_a}$.  
The condition (\ref{cond_f}) means that both wave packets move with the 
same speed and therefore describe a rigid configuration. 
Since our NR ansatz did not include  non-trivial time  dependence 
(apart from linear combination of $\tau$ with $\sigma$ and angular frequency  phases) 
it can only describe such rigid configurations and not those where the magnons move with 
respect to each other.  

 Finally,  we can plot the form of the 
 solutions to understand their behavior. 
In Figs. 1a,1b,1c 
we present  the solutions $\rr_a(\xi)$ for different values of the parameters. 
Notice that $\rr_{2,3}$ are
the densities of $J_{2,3}$ momenta,  so the bumps represent the 
positions of the magnons. 
It can be seen from these  figures that the magnons can 
be separated as much as we want by tuning a parameter. 
Besides the parameters $\omega_a^2$ and $C_a$
 there is a parameter $\labm$ 
that can be loosely associated with the distance
 between the magnons. 
 Notice that none of the conserved quantities depend on $\labm$.

\FIGURE[ht]{
\begin{picture}(170,140)
\put(0,70){\epsfig{file=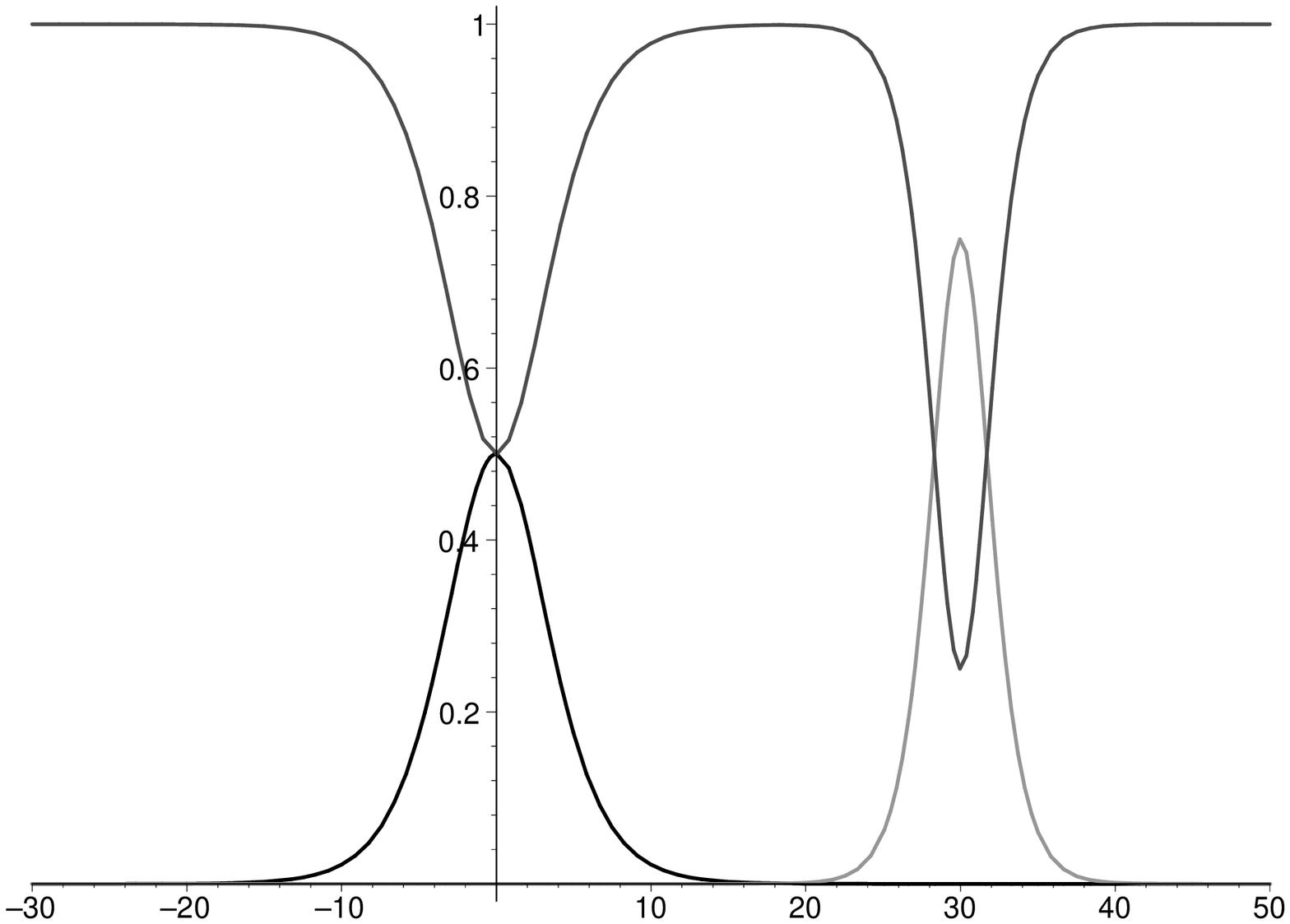, width=85 mm }} 
\put(40,70){1a}
\put(90,70){\epsfig{file=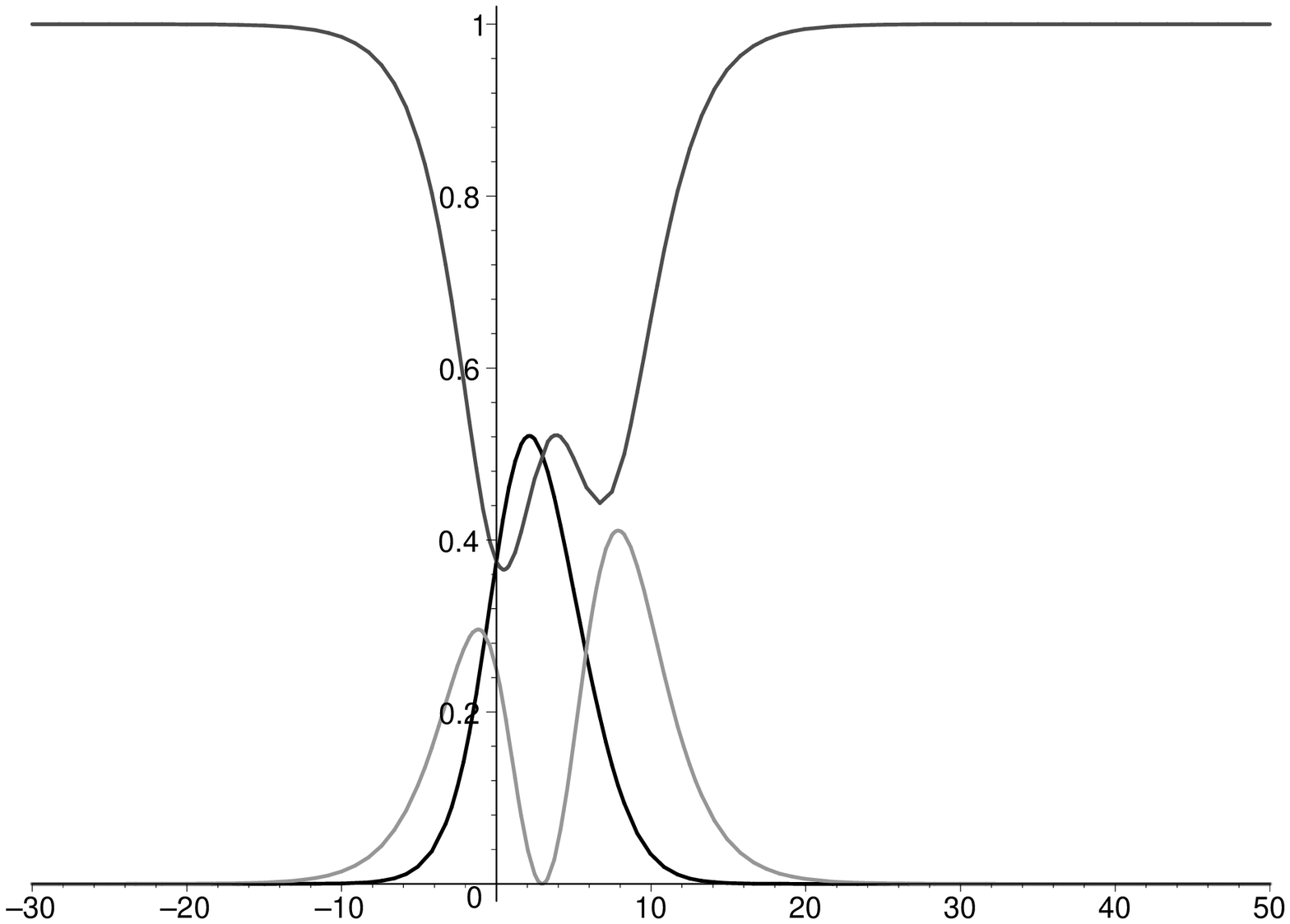, width=85 mm}}  
\put(130,70){1b}
\put(45,0){\epsfig{file=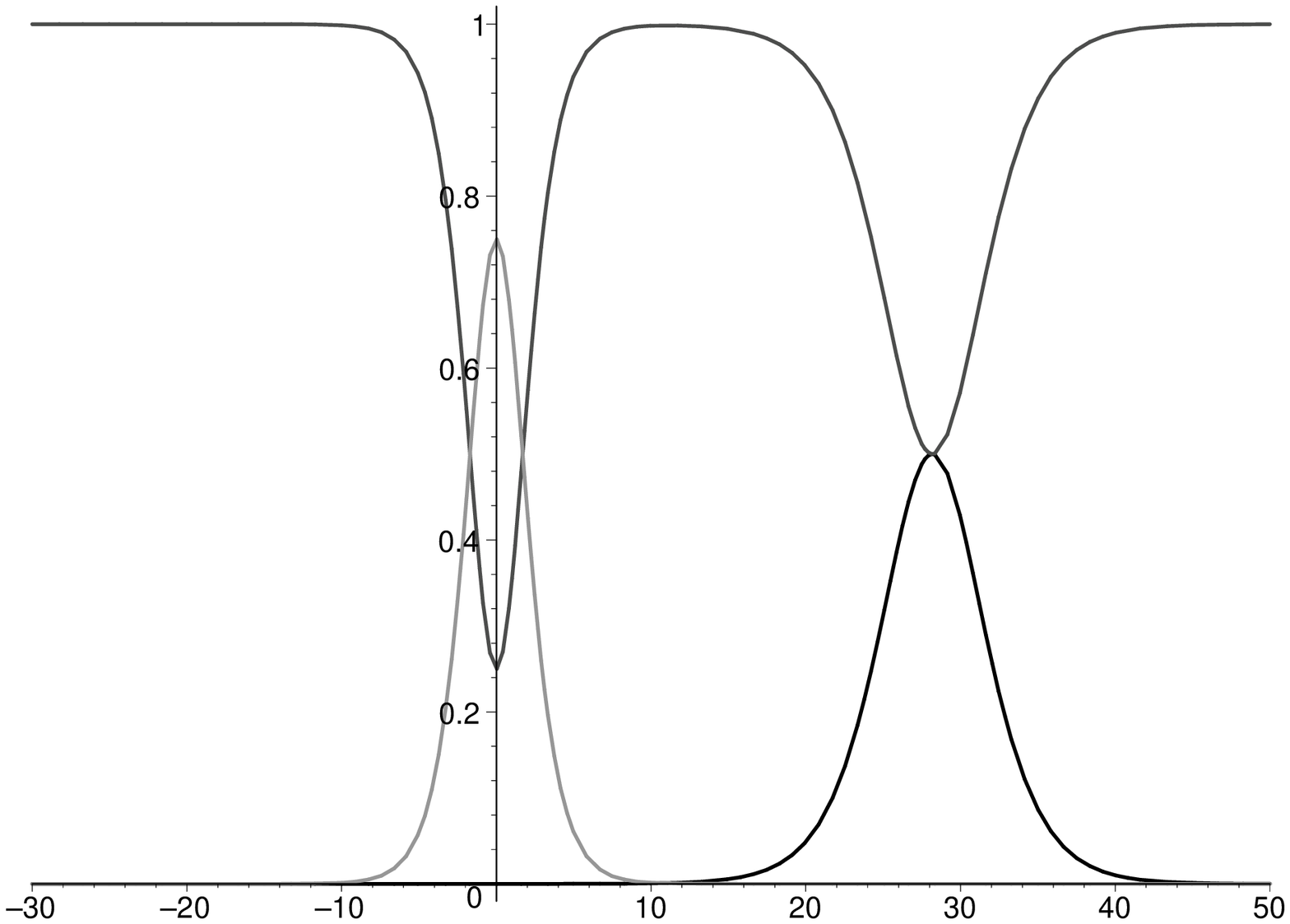, width=85 mm}}
\put(85,0){1c}
\end{picture}
\caption{(1a): The radial functions 
$\rr_a^2(\xi)$ for $\omega_1^2=1$, $\omega_2^2=0.6$, $\omega_3^2=0.2$, $\lab=0.8$, $\labm-0.2=10^{-9}$. The 
curve that goes to $1$ at $\xi=\pm\infty$ is $\rr_1$, while 
$\rr_2, \ \rr_3$ are the gray and black curves going to $0$ at 
$\xi=\pm\infty$. The bumps represent
a concentration of $J_2$ and $J_3$ respectively. 
(1b): Same but with $\labm=0.4$. We see that the bumps moved with
respect to each other. (1c): Same but with $\labm-0.6=-10^{-5}$. 
Comparing to (1a), we see that the positions of the bumps 
interchanged. This occurs as the parameter $\labm$ varies between its limits: $\omega_3^2<\labm<\omega_2^2$.}
\label{fig:sol1}
}

\subsection{Special cases}

Let us consider first the particular case $J_3=0$, $\phi_3=0$. As was pointed out
above, in the  case of $\phi_3=0$ there is no constraint. Now 
the string moves in the $S^3$ part of $S^5$   and the energy formula (\ref{eeefor}) reduces to 
the 2-spin one \cite{Dorey2, MTT}
$$
E-J_1= \sqrt {J_2^2+  {\lambda\over\pi^2 } \sin^2\phi_2}
$$
 and 
reproduced in section 3 using   the present  formalism.



Another  interesting particular case is 
 $J_3=0$,  $\phi_3\neq 0$.  Here  the string
moves  on $S^5$: all $\rr_{1,2,3}$ are non-trivial.
The solution has $w_3=0$ or equivalently $\gamma_3=0$.
Now the energy formula (\ref{eeefor}) reads
\beq
E-J_1= \sqrt {J_2^2+  {\lambda\over\pi^2 } \sin^2\phi_2} + 
{\sqrt{\lambda}\over \pi }\sin\phi_3
\eeq
The last term represents the energy increase due to the
stretching in $\rr_3$ or $\phi_3$.
The stretching is not a free parameter but is determined by the 
constraint $ \cos\phi_3= \cos\gamma_2\cos\phi_2 $. 

\smallskip

In \cite{HM} it was pointed out that a single-spin 
 spinning folded string rotating in $S^2$
 considered in \cite{GKP}, in the limit when 
the ends approach the equator, can be interpreted as a
superposition of two magnons.
The analog solution for $S^5$ can be obtained from our three spin
solution
by setting $\beta =0 $ and  $C_1=0$.
Then $s_1=0$,  $\phi_2=\phi_3={\pi\ov 2}$, and we get the following
energy formula:
\beq
E-J_1= \sqrt {J_2^2+  {\lambda\over\pi^2 }} + \sqrt{J_3^2+ {
      \lambda\over \pi^2 }}
\label{pcc}
\eeq
Note that the constraint (\ref{lkj}) between $J_2$ and $J_3$ 
is absent, because 
$\phi_2=\phi_3={\pi\over 2}$ already solves (\ref{jjhh}).
In the particular case $J_2=J_3=0$, one recovers the expression for the energy of two giant magnons
\beq
E-J_1= 2{ \sqrt {\lambda}\over \pi}
\eeq

\subsection{Large $J_1$ limit of 3-spin circular solution 
}

Finally, it is also interesting to compare the energy of the
above  three spin solution  with the large $J_1$ limit of 
the rigid  circular solution with three angular momenta $J_1,J_2,J_3$ 
found in \cite{ART}.
A similar limit for the two-spin case was considered  in \cite{MTT}.
The energy formula is given by
\beq
E^2=2\sum^3 _a\sqrt{\lambda m_a^2+\nu^2}\ J_a -\nu^2 \ ,\qquad
\sum_a m_a J_a =0\ ,
\eeq
where $\nu $ is determined from 
\beq
\sum_a {J_a\over \sqrt{\lambda m_a^2 +\nu^2}} =1
\eeq
To take the limit of $J_1$
large at fixed $J_2, \ J_3$, we write
$m_2 = n_2 m,\ J_3= n_3 m, \ m_1=-n_1$, and take the limit of large
$m$ with  $n_a$ fixed. The resulting formula is
\beq
E-J_1 ={ 1\over J_1} \Big( J_2 \sqrt{\lambda m_2^2+J_1^2} + J_3
\sqrt{\lambda m_3^2+J_1^2} \Big)
\eeq
with the relation $J_1 m_1+J_2 m_2+J_3 m_3=0$.
In the particular $J_3=0$ case, it reduces to the expression found   in 
\cite{MTT}. Since we are taking the limit of large 
$J_1$ and large $m_2,\  m_3$ with fixed ratio,  ${m_a J_a\over J_1}\equiv k_a$,
the energy formula can be more conveniently written  as
\beq
E-J_1 =  \sqrt{J_2^2 +\lambda  k_2^2  } + 
\sqrt{J_3^2 + \lambda  k_3^2  }\ ,
\eeq
with $ m_1 + k_2 + k_3=0$. The structure is thus similar to the above 
energy formula for the three-spin magnon.

\medskip

One can also consider the same limit for the general circular solution
with spins also on the $AdS_5$ space, i.e. with quantum numbers
$(S_1,S_2,J_1,J_2,J_3)$ and windings $(q_1,q_2,m_1,m_2,m_3)$
(this will generalize the discussion in  \cite{MTT} where  the 
case  of $(S_1,J_1)$ solution was considered).
The energy formula is determined from the equations \cite{ART}
\def \ss {\sum_{a=1}^3 }
\bea\label{kqp}
 \ss {J_a \over  \sqrt{\lambda m^2_a + \nu^2} } =1 \  , \ \ \ \ \ \
\ \ \ \ {E\over\kappa } - 
 \sum_{i=1}^2 {S_i \over \sqrt{\lambda  q^2_i + \k^2}} = 1 \ , \\
\label{oiy}
 2\k E -  2 \sum^2_{i=1} \sqrt{\lambda  q^2_i + \k^2 }\ 
 S_i  - \k^2   =  2 \ss  \sqrt{\lambda m^2_a + \nu^2}\   J_a  - \nu^2 \ ,
 \\
 \label{seo}
\sum_{i=1}^2   q_i S_i  + \ss m_a J_a =0\ . 
\eea
To take the large $J_1$, we make a similar rescaling of the variables as above and, in
addition,
we define $q_i=m p_i$. Then we expand at large $m$ with the new
variables fixed. We find the formula
\beq
E-J_1 ={ 1\over J_1} \Big( J_2 \sqrt{\lambda m_2^2+J_1^2} + J_3
\sqrt{\lambda m_3^2+J_1^2}+S_1 \sqrt{\lambda q_1^2+J_1^2}+ 
S_2 \sqrt{\lambda q_2^2+J_1^2} \Big)
\eeq
or 
\beq
E-J_1 =  \sqrt{J_2^2 +\lambda  k_2^2  } + 
\sqrt{J_3^2 + \lambda  k_3^2  } +  \sqrt{S_1^2 + \lambda  l_1^2  }
+  \sqrt{S_2^2 + \lambda  l_2^2  }\ ,
\eeq
\beq
m_1 + k_2 + k_3+l_1+l_2=0\ .
\eeq
with $k_a\equiv {m_a J_a\over J_1}$, $l_i\equiv {q_i S_i\over J_1}$. 
The expression may be interpreted as the energy 
of a superposition of four bound states of magnons.

\section{Gauge theory (spin chain) interpretation
of rotating giant magnons}

 In the limit $\thc\rightarrow 0$ the theory in question 
  is better described in terms of a perturbative conformal field theory (\N{4} SYM). 
The string corresponds to a field theory operator whose conformal dimension equals the energy of the string. As was shown in \cite{MZ}
in the present scalar operator 
context (and in \cite{SC_QCD} in the context of QCD), 
a useful description of the field theory operators at weak coupling 
is in terms of 
spin chains. In the three spin case we expect the perturbative
description to correspond to an $SU(3)$ spin chain corresponding 
to operators made out of the fields $X=\Phi_1+i \Phi_2$, 
$Y=\Phi_3+i\Phi_4$ $Z=\Phi_5+i\Phi_6$.\footnote{Since we are interested 
in  the limit $J_1\rightarrow \infty$ while 
keeping $J_{2,3}$ finite, 
we are effectively breaking the symmetry
 from $SU(3)$ to $U(1)\times SU(2)$. The $SU(2)$ subgroup rotates the
fields $Y$ and  $Z$ and can be used to
classify  the states.}

 Before going into the details of the spin chain 
  description,  let us note  that a naive extrapolation of the results we already have
from the string side would give, in the $\thc\rightarrow0$ limit:
\beqa
&&\Delta = J_2+J_3 + \frac{\thc}{2\pi^2 J_2} \sin^2\phi_2 + \frac{\thc}{2\pi^2 J_3} \sin^2\phi_3  \\
 &&\frac{J_2}{\sin(2\phi_2)} =  \frac{J_3}{\sin(2\phi_3)}  \label{cond_0}\ . 
\eeqa
 Setting $\phi_a=2p_a$, this expression is the same as  the energy 
of two magnons of momenta $p_2$ and $p_3$, each 
being a bound state of, respectively,  $J_2$ and $J_3$ elementary excitations or ``particles''. The ``particle'' making up
the magnon with momentum 
$p_2$ is actually the field $Y$ 
 and the magnon with momentum 
$p_3$ --  the field $Z$    (each 
inserted into the infinite chain of fields $X$). 
The operator in question should then have 
$J_2$ of $Y$'s, $J_3$ of $Z$'s and an infinite number of $X$'s.\footnote{Note that in $\Delta$ 
we replace $J_2+J_3$ of $X$'s by  $J_2$ of $Y$'s and $J_3$ of  $Z$'s
 and therefore $\Delta=E-J_1$ has a zero order contribution 
of $J_2+J_3$ which is the variation in $J_1$.}   
 
 Given  that the system is integrable, we expect
that both 
the energies and the momenta of the two magnons superpose,
\beqa
 p &=& p_2+p_3  \ \ \Rightarrow  \ \ \ p=\half \dmu  \ , 
\eeqa
 i.e. we 
  also find the relation $p=\half \dmu$ for the total momentum of
   the configuration \cite{HM}.
   
 The classical string configurations  should actually represent a coherent superposition of magnons localized in two wave packets. 
The condition (\ref{cond_f})  or its $\thc\rightarrow 0$ limit (\ref{cond_0}),
 means that the wave packets move at the same speed and therefore the
configuration is rigid. This is because the velocity of the wave
packet is 
the group velocity $v=\frac{\partial \Delta(p)}{\partial p}$. 

 Thus  at $\thc\rightarrow 0$ we  reproduce the main features
 of the three spin magnon configuration 
in a straightforward manner. The result for
all $\thc$ of course follows if we assume that the exact all-loop  magnon energy
 is given as in \cite{Dorey,Dorey2}
  by    $\Delta = \sqrt{J^2+ \frac{\thc}{\pi^2} \sin^2 \frac{p}{2}}$ 
  and again use superposition and the condition of equal velocity.


 \subsection{Bethe ansatz wave function}

 We want to construct the wave function of two magnons, each of them 
being a bound state of several excitations. Again, we start with  
 an infinite chain of
sites with fields $X$
 in which we replace $J_2$ of $X$'s  by $Y$'s 
 and $J_3$ of $X$'s  by $Z$'s. 
 The one-loop 
 $SU(3)$ spin chain Hamiltonian, whose spectrum describes 
 the possible configurations, is 
given by~\cite{MZ}
\beq
 H = \frac{\thc}{8\pi^2} \sum_l \left(1 - P_{l,l+1}\right)\ ,
\eeq
where $P_{l,l+1}$ permutes the sites $l$ and $l+1$. 

 Here we are interested 
  in the case of an infinite spin 
  chain with a finite number of particles (excitations).
 The case of a finite density of particles, namely the 
 thermodynamic limit in  the $SU(3)$ sector, was considered
  in~\cite{BA_SU3}.
 This was done to interpret, in the field theory, the
  string solutions found in~\cite{FT,ART}. 
   In that case one can also use coherent state
methods to compare directly the actions  for relevant low-energy modes on 
 the string and the spin chain side~\cite{HL,ST}.

We shall follow closely the ideas in \cite{Yang} and \cite{Suth}.
It is important to give a detailed description of the problem in order
to get a precise idea of which states exist,  so that we can identify the 
 string solution
 found above  with an operator 
 on the field theory side. 
 To start with the Bethe ansatz let us assume that we 
 add $N=J_2+J_3$ distinguishable particles and later 
symmetrize as appropriate. 
The configurations are divided into
 sectors labeled by a permutation $Q=(Q_1,\ldots, Q_N)$,  where $Q_i$ are integers 
from $1$ to $N$ which  are all different. 
$Q_1$ is the left-most particle, $Q_2$ the next one and $Q_N$ the
right-most 
one. For example,  $Q=(3,1,2)$ 
means that we put the third particle on the left, the first one 
in the middle and the second one on the right (recall 
that they are distinguishable for now).
 Then we take $N$ momenta $k_i$ all different
  and assign them to each particle according 
  to another permutation $P=(P_1,\ldots, P_N)$. 
This means that $k_{P_1}$ is the momentum of 
the first particle and so on. 
The Bethe ansatz gives a wave function in each sector labeled by $Q$ as:
\beq
 \psi_Q(x_1,\ldots, x_N) = \sum_P A(Q|P)\  e^{i(k_{P_1} x_{Q_1} + 
 \dots + k_{P_N} x_{Q_N})}\ , 
\eeq
 where $x_n$ is an integer which describes the position of the $n$-th particle. Notice that $x_{Q_1}< \ldots <x_{Q_N}$. There are $(N!)^2$ 
coefficients $A(Q|P)$ that we need to determine from the condition that 
$\psi$ is an eigenstate of the above Hamiltonian. 

\def \no {\nonumber}

When the particles are far apart, applying $H$, we find that the energy is given by
\beq
 E = \frac{\thc}{2\pi^2} \sum_{l=1}^N \sin^2 \frac{k_l}{2}\ .
\eeq
 When two particles, e.g., $Q_l$ and $Q_{l+1}$, 
   come together (meaning that $x_{Q_l}=x_{Q_{l+1}}\pm 1$)
    the eigenstate condition determines that
\beqa
\lefteqn{ \ekplo A(\tilde{Q}|P) + \ekpl A(\tilde{Q}|P')  } \ \ \ \ 
 && \no \\
      &=& -\left(\ekplo-\ekpl\ekplo-1\right)A(Q|P) - 
      \left(\ekpl-\ekpl\ekplo-1\right) A(Q|P')\ , \ \   \\  
\lefteqn{ \ekplo A(Q|P') + \ekpl A(Q|P')   } \ \ \ \  &&\no  \\
    &=& -\left(\ekplo-\ekpl\ekplo-1\right)A(\tilde{Q}|P)
     - \left(\ekpl-\ekpl\ekplo-1\right) A(\tilde{Q}|P')\ ,   
\eeqa
 where $\tilde{Q}=(Q_1,\ldots,Q_{l+1},Q_l,\ldots,Q_N)$, 
 namely, the same as $Q$ but  with two particles interchanged.
  The same applies to
$P'=(P_1,\ldots,P_{l+1},P_l,\ldots,Q_N)$ but
 now we interchange the momenta we 
 assign to the two particles. We can solve for $A(Q|P')$ as
\beq
 A(Q|P') = \alpha_{P_l,P_{l+1}} A(Q|P) + \beta_{P_l,P_{l+1}} A(\tilde{Q}|P),
  \ \ \ \alpha_{ij} = \frac{i}{u_i-u_j+i},\ \ \  \ \ \beta_{ij}=\alpha_{ij}-1\ , 
\eeq
 where we defined
\beq
 u_i = \half \cot \frac{k_i}{2}\ .
\eeq
 The way to solve these equations is to assume first that we know $A(Q,\one)$ (where $\one=(1,2,\ldots,N)$ is the identity permutation) and compute
$A(Q|P)$ for all $P$. Notice that in principle we  only know  how to do
 permutations that interchange two consecutive momenta, but it is easy to see
that in this way  we can get to an arbitrary permutation. 
If we define a set of $N!$ vectors $\xi_P$ as the columns 
of $A$ (\ie\ $(\xi_P)_Q = A(Q|P)$)\footnote{It is conventional to call this vector $\xi_P$. Of course it bears no relation to the 
world-sheet coordinate $\xi$ we used in previous sections.}
we get
\beq
  \xi_{P'} = \left(  \alpha_{P_l,P_{l+1}}  + 
  \beta_{P_l,P_{l+1}} \hat{P}_{l,l+1} \right) \xi_P = Y_{l,l+1} \xi_P\ , 
\eeq
 where $\hat{P}_{l,l+1}$ is an operator that 
 interchanges the components of $\xi_P$ such that $(\hat{P}_{l,l+1}\xi_P)_Q 
 = (\xi_P)_{\tilde Q}$.
 
 As was mentioned above, given $\xi_\one$ we can construct $\xi_P$ for all
 $P$. However, 
this construction works provided
 certain compatibility conditions hold. 
One is that if we do a permutation twice we should get 
the identity $(P')'=P$. The other stems from the fact that, for example, we can interchange
the first and third momenta in two different ways which have to agree:
 $Y_{13}= Y_{12} Y_{23} Y_{12} = Y_{23} Y_{12} Y_{23}$. These are the
Yang-Baxter conditions that here read
\beqa
\alpha_{21}\alpha_{12} + \beta_{12} \beta_{21} &=& 1 \\
\beta_{21} \alpha_{12} + \alpha_{21}\beta_{12} &=& 0 \\
\alpha_{13}\alpha_{23}\beta_{12} + \alpha_{13}\alpha_{12}\beta_{23} - \alpha_{12} \alpha_{23}\beta_{13} &=& 0
\eeqa 
 and can be easily checked.

 If we want a scattering state,  we are done: 
we have to specify an arbitrary $\xi_\one$ and that is it. 
If some of the particles are indistinguishable
we need to impose symmetry conditions on $\xi_\one$. 
For example, if they are all of the same type, we have to take all components 
of $\xi_\one$ equal: $(\xi_\one)_Q=1$ for all $Q$ and so on.

 If we want the state to be that of a periodic chain then 
 we have to impose periodicity conditions which are 
 non-trivial and require what amounts to
another Bethe ansatz for the components of $\xi_\one$. 
This is the nested Bethe ansatz that results in the
 Bethe equations that, as we already mentioned 
were discussed in this context in~\cite{BA_SU3}.

 If we want to find bound states on an  infinite chain,
 which is our main interest here, 
  we have to impose certain conditions on $\xi_\one$ 
  that we are going to study below.
Before doing that in general  we are
 going to work out the examples of two and three particles.  

\subsection{Two particle states}

 If there are two particles we have two permutations that we can
  call $\one=(12)$ and $\two=(21)$. Therefore, 
   there are two vectors $\xi_\one$, $\xi_\two$
of two components each. We get:
\beq
 \xi_\one=\left(\begin{array}{c}a\\b\end{array}\right), \ \ \ \hat{P}_{12}\xi_\one=\left(\begin{array}{c}b\\a\end{array}\right) \ \ \ \Rightarrow \ \ \ 
\xi_\two=\left(\begin{array}{c}\alpha_{12}a+\beta_{12}b
\\\beta_{12}a+\alpha_{12}b\end{array}\right).
\eeq
 Suppose now that $\Ima(k_1) <0$ and $\Ima(k_2) >0$. We get a bound state
  if we assign $k_1$ to the particle to the left and $k_2$ to the right. If
we interchange the momenta we get a wave function that diverges at $\pm\infty$. Therefore,
 we should have 
$\xi_2=0$.  This gives equations for $a$ and $b$ that are compatible only if $\alpha_{12}\pm\beta_{12}=0$. Since $\alpha_{12}-\beta_{12}=1$ we can
only have $\alpha_{12}+\beta_{12}=0$. Then  $a=b$,
namely,  there is a bound state in the symmetric sector. Furthermore, 
\beq
 \alpha_{12} + \beta_{12} = 0  \ \ \Rightarrow\ \ \ u_1-u_2=i
\eeq
 Since the total momentum and the energy are real, 
  we need $k_1=k_2^*$ which implies $u_1=u_2^*$. The solution is
\beq
 u_1 = \buo +\frac{i}{2}, \ \ \ \ \ \ \  u_2 = \buo - \frac{i}{2}\ . 
\eeq
 The total momentum and energy are
\beqa
 p &=& k_1 + k_2 = 2\,\Rea(k_1) \\
 E &=& \frac{\thc}{2\pi^2}\sin^2 \frac{k_1}{2} +
  \frac{\thc}{2\pi^2}\sin^2 \frac{k_2}{2} = \frac{\thc}{4\pi^2}\sin^2 \frac{p}{2}\ . 
\eeqa
 The (not normalized) wave function is 
\beq
 \ket{\psi}(y_1,y_2) = \left[ \ket{YZ}+\ket{ZY} \right]
  e^{i\Rea(k_1)(y_1+y_2)} e^{-\Ima(k_1) (y_1-y_2)}\ , 
\eeq
 where we defined $y_i=x_{Q_i}$ so that $y_1$ is the position of the 
 particle at the left and $y_2$ the position of that 
at the right (\ie\ $y_1<y_2$, also $\Ima(k_1) <0$). Also, we used
a ket notation for the vector $\xi$. The state $\ket{YZ}$
 means that the particle on the left is a $Y$ and that on the right a $Z$. 
 The opposite applies to 
$\ket{ZY}$. If both particles are $Y$ then we simply get 
\beq
 \ket{\psi}(y_1,y_2) = \ket{YY}\ e^{i\Rea(k_1)(y_1+y_2)} e^{-\Ima(k_1) (y_2-y_1)}\ . 
\eeq

\subsection{Three particle states}

  Now there are six permutations that we can label as:
\beq
\one=(123), \ \two=(132),\ \three=(312),\ \four=(213),\ \five=(231),\ \six=(321)
\eeq
 Thus, $\xi_P$ is a six-vector. Recall  that the different components of $\xi_P$ correspond to different orderings of the particles and 
the different vectors $\xi_P$ to different momenta assignments. On $\xi_\one$
 the permutations act as:
\beq
 \xi_\one = \left(\begin{array}{c} a\\b\\c\\d\\e\\f \end{array} \right), 
\ \ \hat{P}_{12}\xi_\one=\left(\begin{array}{c} d\\c\\b\\a\\f\\e \end{array} \right), 
\ \ \hat{P}_{23}\xi_\one= \left(\begin{array}{c} b\\a\\f\\e\\d\\c \end{array} \right)
\eeq
 This follows,  for example,  from the fact that $\hat{P}_{12}$
 interchanges 
$\one\leftrightarrow\four$, $\two\leftrightarrow\three$, $\five
\leftrightarrow\six$ and similarly for  $\hat{P}_{23}$. 

 For a bound state with real energy and momentum, let us consider $\Ima(k_1) <0$, $\Ima(k_2)=0$, $-\Ima(k_1) = \Ima(k_3)>0$.  It is clear that 
we can have a bound state only if $\xi_{P}=0$ for $P\neq \one$, \ie\ the only possibility is that $k_1$ goes to the left, $k_2$ in 
the middle and $k_3$ to the right. For this we only need to require  that 
\beqa
 \xi_\two = \left[\alpha_{23}+\beta_{23}\hat{P}_{23}\right]\xi_\one = 0 \ , \ \ \ \ \ \ \ \ \ \ 
 \xi_\four = \left[\alpha_{12}+\beta_{12}\hat{P}_{12}\right]\xi_\one = 0 \ . 
\eeqa
 There is a non-vanishing solution for $\xi_\one$ only if $\alpha_{12}=-\beta_{12}$ and $\alpha_{23}=-\beta_{23}$ which is equivalent to
\beq
 u_2-u_3=u_1-u_2 =i  \ \ \ \Rightarrow\ \ \ \  u_1=\buo+i,\ \ \ u_2=\buo,\ \ 
  \ u_3=\buo-i \ \ \  (\buo\ \mbox{is real})
\eeq
 Given those values of momenta we see that the solution is such that
 $a=b=c=d=e=f$, namely it is in the totally symmetric sector. 
 The energy and momentum are:
\beqa
 p&=& k_1+k_2+k_3  \ \ \Rightarrow \ \ \tan\frac{p}{2}=\frac{3}{2\buo}, \\
 E&=& \frac{\thc}{2\pi^2}\sum_{l=1}^3 \sin^2\frac{k_l}{2} = \frac{\thc}{6\pi^2} \sin^2\frac{p}{2}
\eeqa
 If there are two particles of type $Z$ and one of type $Y$ the wave function is 
\beq
\ket{\psi}(y_1,y_2,y_3) = \left[\ket{YZZ}+\ket{ZYZ}+\ket{ZZY}\right] e^{i(k_1y_1+k_2y_2+k_3y_3)}
\eeq
 A natural question is if there are states (in the other symmetry sector) 
 which describe scattering of a single particle and a two-particle bound state.
For that we choose $\Ima(k_1) =0$, $\Ima(k_2)<0$, $\Ima(k_3)>0$ and 
consider permutations such that $k_2$ is always to  the left of $k_3$ so that the wave
function does not diverge. It is clear that we only have to kill $\xi_\two$. Namely,
 $\alpha_{23}=-\beta_{23}=\half$, $u_2-u_3=i$. If the reference
configuration is $\ket{YZZ}$,  then we find from 
the symmetry that $a=b$, $c=d$ and $e=f$
 since they multiply the same configuration.  This means that there
are three independent states that we can choose to be
\beqa
 \ket{1} &=& \sqrt{\frac{2}{3}} \left[ \ket{YYZ} -{\textstyle \half} \ket{ZYZ} -{\textstyle \half}
  \ket{ZZY} \right] = \ket{\half \half} \\
 \ket{2} &=& \frac{1}{\sqrt{2}} \left[ \ket{ZYZ} - \ket{ZZY} \right] = \ket{\half\half}'\\
 \ket{3} &=& \frac{1}{\sqrt{3}} \left[ \ket{YYZ} + \ket{ZYZ} + \ket{ZZY} \right]=\ket{\frac{3}{2}\half} 
\eeqa
 We used also an alternative notation in terms of spin $\half$ representations by identifying $Y$ with spin down and $Z$ with spin up.
The last state $\ket{3}$ is in the symmetric sector and we ignore it. 
If we apply the condition $\xi_\two=0$,  we need again $u_3-u_2=i$ but also
$c=f$ which means that the state is $\xi_\one=\ket{1}$. 
The other non-vanishing vectors are $\xi_\four$ and $\xi_\five$ which can be computed  from 
\beq
\xi_\four = \left[\alpha_{12} + \beta_{12}\hat{P}_{12}\right] \xi_\one,\
 \ \ \   \ \ \ \xi_\five = \left[\alpha_{13}+\beta_{13} \hat{P}_{23}\right]\xi_\four
\eeq
Finally,  we obtain the wave function
\beqa
 \ket{\psi}(y_1,y_2,y_3) &=&   \sqrt{\frac{2}{3}}             \left[\ket{YZZ}-\half \left(\ket{ZYZ}+\ket{ZZY}\right)\right] 
                  e^{i(k_1y_1+k_2y_2+k_3y_3)} \no \\
                         & & - \sqrt{\frac{2}{3}} \beta_{12}  \left[\ket{ZZY}-\half \left(\ket{YZZ}+\ket{ZYZ}\right)\right] 
                  e^{i(k_2y_1+k_3y_2+k_1y_3) }\no \\
                         & & - \sqrt{\frac{2}{3}}             \left[\ket{ZYZ}-\half \left(\ket{YZZ}+\ket{ZZY}\right)\right] 
                  e^{i(k_2y_1+k_1y_2+k_3y_3)}\no \\
                         & & - \sqrt{\frac{2}{3}} \alpha_{12} \left[\ket{ZZY}-\half \left(\ket{YZZ}+\ket{ZYZ}\right)\right] 
                  e^{i(k_2y_1+k_1y_2+k_3y_3)} 
\eeqa  
 We see that if $y_1\rightarrow -\infty$ only the first line survives (since $\Ima(k_2)<0$) and it precisely describes a particle on the left and two 
symmetrized particles on the right,
 as we expect for a particle moving away from a two particle bound state. 
Similarly, if $y_3\rightarrow\infty$ only the second line survives 
describing a bound state to the left and a single particle to the right. 

 It is clear also that we do not see any bound state (of the three 
particles) in this sector. This suggests that the string solution that we are considering should correspond to a state of two magnons which 
are not bound to each other. To describe such a state we shall first 
 review the construction that gives one bound state and then  extend it to two magnon case.  

\subsection{$J$-particle bound state}

 The bound state of $J$ particles is in the
  symmetric sector and was found already by Bethe
   in his original paper \cite{Bethe}. Here we review briefly 
this  construction since these bound states are the field theory analog of the 
giant magnon with an extra angular momentum~\cite{Dorey,MTT}. Again,  we choose the momenta such 
that only $\xi_\one\neq 0$. For this to happen permuting any 
successive momenta should give zero, 
 which implies that $u_{j+1}-u_j=i$ and all components of $\xi_\one$ 
 are equal, namely the symmetric sector. 
 Again, taking into account that the energy and momenta should be real,  we obtain:
\beq
 u_j = \buo - \frac{J-1}{2}i + j\, i, \ \ \ \ \ \ \ \ \ \ j=0, \ldots,  J-1\ , \ \ (\buo\ \mbox{is real}). 
\eeq
 Using that
\beq
 u_j = \half\, \cot\frac{k_j}{2} \ \ \ \Rightarrow \ \ \ e^{ik_j} = \frac{u_j+\frac{i}{2}}{u_j-\frac{i}{2}} 
\eeq
 and defining 
\beq
a_j = u_j-\frac{i}{2} = \buo - \frac{J}{2} i + j\, i
\eeq
 we have for the total momentum
\beq
 e^{ip} = e^{i\sum_{j=0}^{J-1}k_j} = \prod_{j=0}^{J-1} \frac{u_j+\frac{i}{2}}{u_j-\frac{i}{2}} = \prod_{j=0}^{J-1} \frac{a_{j+1}}{a_j} 
 = \frac{a_J}{a_0} = \frac{\buo+\frac{J}{2}i}{\buo-\frac{J}{2}i}\ , 
\eeq
 Thus 
\beq
 \tan\frac{p}{2} = \tan \phi = \frac{J}{2\buo}\ , 
\eeq
 where we used the notation $\phi=\frac{p}{2}$ as in the  previous
 sections. 
This exhibits the fact  that, in the $u$-plane, the angle $\phi$ has a
simple interpretation, as illustrated 
 in Fig. \ref{fig:2mag}, where two magnons are shown. 

 The resulting expression for the energy  is 
\beqa
E &=& \frac{\thc}{2\pi^2} \sum_{j=0}^{J-1} \sin^2\frac{k_j}{2} = \frac{\thc}{8\pi^2} \sum_{j=0}^{J-1} \left(2-\frac{a_{j+1}}{a_j}-\frac{a_j}{a_{j+1}}\right) \\
  &=& \frac{\thc}{8\pi^2}\left(2-\frac{a_1}{a_0}-\frac{a_{J-1}}{a_J}\right)  = \frac{\thc}{8\pi^2}\frac{4J}{J^2+4\buo^2} 
      = \frac{\thc}{2\pi^2 J} \sin^2\frac{p}{2}
\eeqa
where we used that
\beq
 a_{j-1}-2a_j+a_{j+1}=0\ \ \ \ \Rightarrow \ \ \ \ \frac{a_{j-1}+a_{j+1}}{a_j}=2
\eeq
 to simplify the sum. We see that the state is indeed a {\it  bound state}
  since the total energy is less than the energy of $J$ particles of 
momentum $\frac{p}{J}$
\beq
 E = \frac{\thc}{2\pi^2J}\sin^2\frac{p}{2}\ \ \le \ \ \frac{\thc}{2\pi^2} J \sin^2\frac{p}{2J}
\eeq
 For $p\rightarrow 0$ the binding energy goes to zero;
   therefore,  at small momentum, such bound states can be ignored.  
   
   The relation between Bethe bound states  of elementary magnons (``Bethe strings'')
    and giant magnons was also pointed out 
   in \ci{MTT} where it was  generalized  to all orders in $\l$   by starting with 
   the asymptotic BDS Bethe ansatz \ci{bds}.

Another feature 
  is that to construct a semi-classical state
 we should superpose magnon states to create a wave packet. As is well known, 
such wave packets move at the group velocity given by
\beq
v = \frac{\partial E}{\partial p} = \frac{\thc}{4\pi^2 J} \sin p 
\eeq
 Again,  there is a nice geometric interpretation. In Fig.\ref{fig:2mag}
  we draw a circle going through the origin and the points
$(\buo,\frac{J}{2})$ and $(\buo,-\frac{J}{2})$. The center of the circle is at a distance $\frac{\thc}{8\pi^2}\frac{1}{v}$ from the origin. 
In the figure both magnons move with the same velocity so
that  the circles coincide.

\subsection{Two-magnon state}

 To reproduce the results from the string side we make the simple ansatz that there are two bound states, one with $J_2$ particles and the other
with $J_3$. We take the initial configuration of momenta as
\beq
 u_1, u_2, \ldots , u_{J_2}, \tilde{u}_1, \tilde{u}_2, \ldots, \tilde{u}_{J_3},
\eeq
 where the $u$'s determine the momenta of the particles in the first bound state and $\tilde{u}$ in the other. 
We now allow permutations such that the order of the $u$'s is preserved and the same for
 the $\tilde{u}$'s. This still allows for 
$\bigg(\begin{array}{c} J_2+J_3 \\ J_2\end{array}\bigg)$ permutations,
 namely,  non-vanishing $\xi_P$ vectors. It is clear that to satisfy this we
only need to require that permutations of successive $u$'s or successive $\tilde{u}$'s vanish which give the standard bound state conditions for 
$u$ and $\tilde{u}$ that we already discussed, namely:
\beq
 u_j = \buo_2 - \frac{J_2-1}{2}\,i + j\, i, \ \ \ j=0,\ldots,  J_2-1;\ \ \ \ \
 \ \ \   \tilde{u}_j = \buo_3 - \frac{J_3-1}{2}\,i + j\, i, \ \ \ j=0,\ldots, J_3-1\ . 
\eeq  
 An example is given in Fig.\ref{fig:2mag}.
The wave function $\xi_\one$ has to be such that it is invariant  under permutations of the first $J_2$ particles and the last $J_3$. This is automatically
satisfied if we choose the state $\ket{\underbrace{Y\ldots Y}_{J_2} \underbrace{Z\ldots Z}_{J_3}}$. However,  this is not in the sector we want. 
If we consider $Y$ to be a spin up and $Z$ to be a spin down we want the state of spin $J_2-J_3$ (and $z$ projection $J_2-J_3$). 
It is clear that the state in question is obtained by symmetrizing the first $J_2$ components and the last $J_3$ ones such that we get two states 
with spins $J_2$ and $J_3$. Then we compose both to total spin $J_2-J_3$. We can 
therefore express it as 
\beq
 \ket{\xi_1} = \sum_{M_2+M_3=J_2-J_3} \left(\begin{array}{ccc}J_2& J_3& J_2-J_3 \\ M_2 & M_3 & -M_2-M_3 \end{array}\right) 
               \ket{\overbrace{\,(\,\underbrace{Y\ldots Y}_{J_2+M_2} \underbrace{Z\ldots Z}_{J_2-M_2}\,)\,}^{\mbox{symmetrized}}\,
                   \overbrace{\,(\,\underbrace{Y\ldots Y}_{J_3+M_3}\underbrace{Z\ldots Z}_{J_3-M_3}\,)}^{\mbox{symmetrized}}}
\eeq
where the parenthesis indicate that the state should be symmetrized over the position of the corresponding $Y$'s and $Z$'s.
Also, we used the 3-$j$ (Clebsch-Gordan) coefficients:
\beqa
\left(\begin{array}{ccc}J_2& J_3& J_2-J_3 \\ M_2 & M_3 & -M_2-M_3 \end{array}\right) &=&    (-1)^{J_2+M_2} 
  \left[\frac{(2J_3)!(2J_2-2J_3)!(J_2+M_2)!(J_2-M_2)!}{(2J_2+1)!(J_3+M_3)!(J_3-M_3)!}\right]^{\half} \nonumber\\
  && \times\left[ \frac{(J_2+M_2)!(J_2-M_2)!}
  {(J_2-J_3+M_2+M_3)!(J_2-J_3-M_2-M_3)!}\right]^{\half} 
\eeqa
This completely characterizes the state. In a similar way, one can
compute the other $\xi_P$ to write down the complete wave function.
 
A physical way to describe this state is in terms of its $SU(2)$ quantum numbers, where $SU(2)$ rotates $Y$ and $Z$. Under that group, one magnon
carries angular momentum $J_2$ and the other $J_3$. Therefore,
 their constituent particles are, internally, in a totally symmetric state. Now, the 
state of the two magnons can have angular momentum from $J_2+J_3$ to
$J_2-J_3$. All these  states are possible  but we are just
interested in the one  with spin $J_2-J_3$.

 Finally, to establish a correspondence 
  with the string theory picture, 
we need, as we already discused, to construct a semiclassical (coherent) state. 
 Then we get a rigid configuration 
 when 
 the group  velocities of the wave packets representing the 
two giant magnons are equal. We see in Fig.\ref{fig:2mag} 
that the circles drawn for the two magnons coincide. 
 
\FIGURE[ht]{\epsfig{file=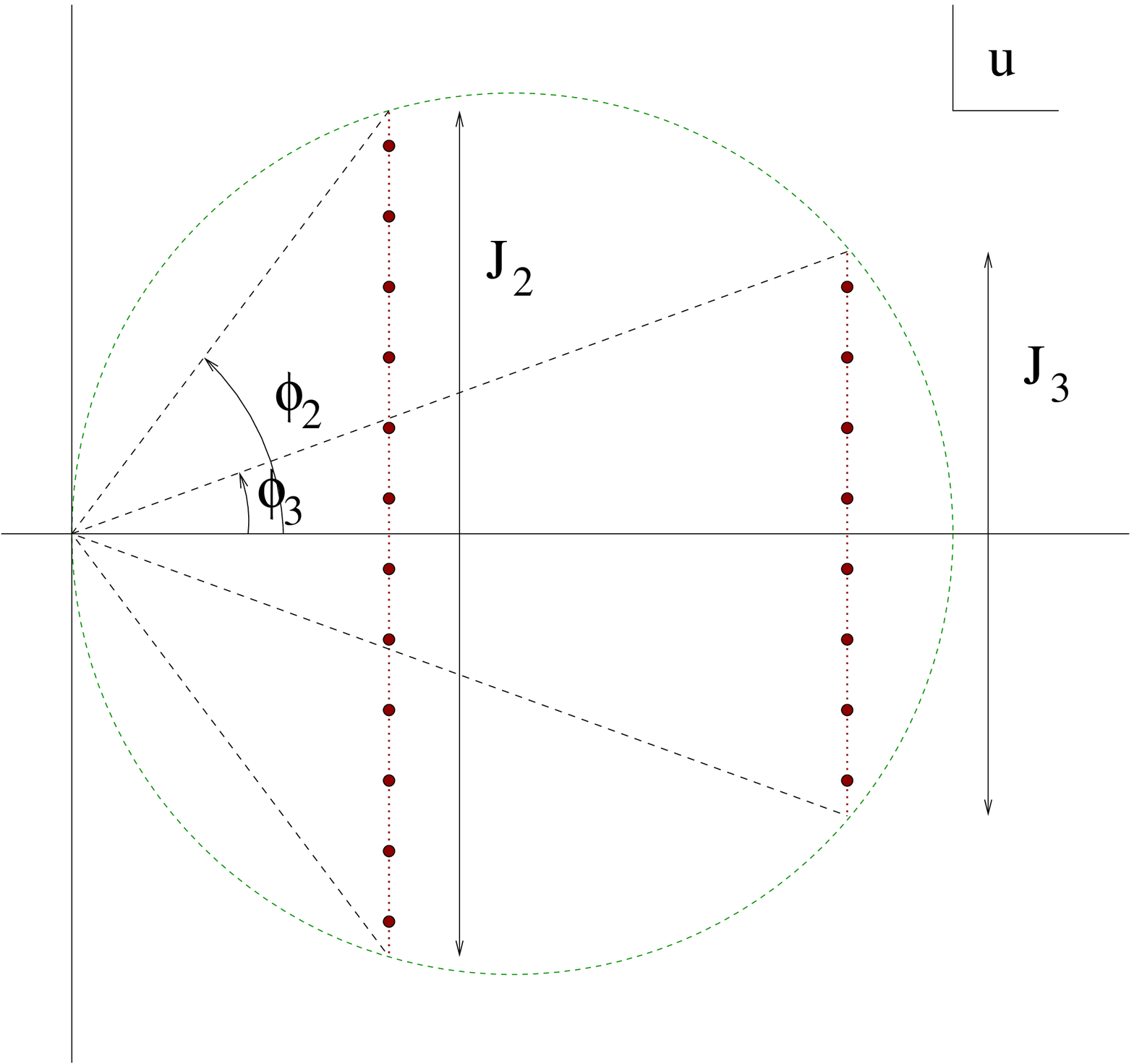, width=12cm}
\caption{Distribution of momenta in terms of $u_j=\half \cot\frac{k_j}{2}$ for the two magnon state. Geometrically,  it is interesting
that the angles shown are half the momenta of 
each magnon and also that the center of the 
circle is at a distance from  the origin equal to the inverse
of the group velocity (which is the same for both 
magnons so there is only one circle).}
\label{fig:2mag}
}

\section{Conclusions} 
 
We have studied a generalized ansatz for strings moving in 
\adss{5}{5} that reduces the problem of finding solutions to that of 
solving the 
Neumann-Rosochatius system. That system describes an effective 
particle moving on a sphere in a specific 
potential. In our case we had an extra term equivalent to a coupling 
to a magnetic field. Such term, however,   appeared only in the equations
for the angular variables. For  the radial coordinates, we still got the usual
NR lagrangian.  After solving the NR system, the trajectory of the 
particle should be understood as the profile of the string. 
Such string rotates rigidly in time according to the ansatz we proposed. 
 
Since the solutions of the Neumann-Rosochatius system are relatively 
simple to find, we extended the 
giant magnon solution to the case of two additional 
angular momenta. Although,  in principle,  the integrability  does not 
guarantee a simple expression for 
the  conserved string quantities (such as angular momenta), we have found  
a rather simple result: the conserved quantities correspond to a 
 superposition of those of two giant magnons, each  carrying 
one 
of the two finite angular momenta. However, since the solution turned out to
describe a rigid string we  got  an 
extra condition that the group velocity of the 
two magnons should be the same. 
It would be interesting to study other solutions (which will no 
longer be described by the NR ansatz) 
where the two magnons move relatively to each other. 
 
In the weak coupling gauge theory limit the description of the 
two magnons is that of two bound states in a spin chain that move freely. 
Here
it is trivial to consider the magnons moving with respect to each other 
since we can see that they do not interact. The wave function 
of such system can be constructed using the Bethe ansatz as we 
discussed in some detail. 
 
An interesting point is that,  on the string side, using the plots we 
presented, 
one can easily differentiate  the two magnons. This suggests that 
one can  directly  relate the position along the spin chain  with  the 
position along the string. It should be interesting to establish a 
more precise map between the action of the string 
and that of the spin chain as can be done at small momentum in the 
``thermodynamic'' limit. 
 
 Finally,  we should note that the ansatz that we used here can be generalized to 
the full \adss{5}{5}  case (as in \cite{ART}); one   can  also 
 include some pulsating solutions by interchanging
the $\sigma$ and $\tau$ world-sheet coordinates. It would be interesting to understand these
other solutions and see if there is an analog of the giant magnon 
solution in those  larger sectors.

\section{Acknowledgments}

 We are very grateful to I. Klebanov, J. Minahan, R. Roiban
   and A. Tirziu for various comments and discussions. 
 M.K.  and A.A.T. are also grateful to the KITP at Santa 
 Barbara  for hospitality during the 2004 
``QCD and Strings'' workshop where some of the  ideas of  this work were
 conceived.

The work of  J.R. and A.A.T.  
is supported in part by the European EC-RTN network MRTN-CT-2004-005104. 
J.R. also acknowledges support by MCYT FPA 2004-04582-C02-01.
The work of A.A.T. is supported in part by the DOE grant DE-FG02-91ER40690, 
 INTAS  03-51-6346  and the RS Wolfson award.
 The work of M.K. was supported in part by the National Science Foundation Grant No. PHY-0243680. 
Any opinions, findings, and conclusions or recommendations expressed in this material are those of 
the authors and do not necessarily reflect the views of the National Science Foundation.

\bigskip

\noindent {\bf Note added}
\medskip

While this paper was in preparation, there appeared  two papers  
  \cite{spradlin} and \cite{BR}  which
also discuss spinning giant magnons on $S^5$.
The three-spin solution presented  at the end of   \cite{spradlin} corresponds to
a special  case of our  solution with energy given by
 (\ref{pcc}) and having  $s_1=0$,
$\phi_2=\phi_3={\pi\over 2}$.
At the same time, we do not understand  the three-spin solution presented
in sect. 2.2 of \cite{BR}.\footnote{In that solution, the condition
 $r_2=r_3$ (or $\psi=\pi/4$ in the notation of
\cite{BR}) is imposed by hand, but this ansatz, with $C_2=C_3=0$ and $w_2\neq
w_3$, does not satisfy the equations
of motion for $r_2$ and $r_3$, see eq. (\ref{raaa}). The equations are
satisfied only in the case $w_2=w_3$, but this is the
two spin solution with energy (\ref{twoJJ}) as can be seen by an orthogonal
transformation.}

\end{document}